\documentclass[]{aastex631}

\newcommand{\pmass}{$2.64$}
\newcommand{\smass}{$2.10$}
\newcommand{\perr}{$ \pm 0.07$}
\newcommand{\serr}{$ \pm 0.03$}

\newcommand{\intradthirtyone}{$8.36 \pm 0.15$}
\newcommand{\lumradthirtyone}{$8.18 \pm 0.13$}
\newcommand{\lumthirtyone}{$63.6 \pm 1.0$}

\newcommand{\primage}{$533$}
\newcommand{\primageerr}{$^{+42}_{-38}$}

\newcommand{\mircx}{\mbox{MIRC-X} }

\shorttitle{Precise Age of 12 Com}
\shortauthors{Lam et al.}
\graphicspath{{./}{figures/}}

\begin{document}

\title{Precise Age for the Binary Star System 12 Com in the Coma Berenices Cluster}

\author{Rex Lam}
\affiliation{San Diego State University, Department of Astronomy, San Diego, CA, 92182 USA}

\author[0000-0003-4070-4881]{Eric L. Sandquist}
\affiliation{San Diego State University, Department of Astronomy, San Diego, CA, 92182 USA}

\author[0000-0001-5415-9189]{Gail H. Schaefer}
\affiliation{ The CHARA Array of Georgia State University, Mount Wilson Observatory, Mount Wilson, CA
13 91023, USA}

\author[0000-0001-9939-2830]{Christopher D. Farrington}
\affiliation{ The CHARA Array of Georgia State University, Mount Wilson Observatory, Mount Wilson, CA
13 91023, USA}

\author[0000-0002-3380-3307]{John D. Monnier}
\affiliation{ The CHARA Array of Georgia State University, Mount Wilson Observatory, Mount Wilson, CA
13 91023, USA}

\author[0000-0002-2208-6541]{Narsireddy Anugu}
\affiliation{ The CHARA Array of Georgia State University, Mount Wilson Observatory, Mount Wilson, CA
13 91023, USA}

\author[0000-0001-9745-5834]{Cyprien Lanthermann}
\affiliation{ The CHARA Array of Georgia State University, Mount Wilson Observatory, Mount Wilson, CA
13 91023, USA}

\author[0000-0002-4313-0169]{Robert Klement}
\affiliation{ The CHARA Array of Georgia State University, Mount Wilson Observatory, Mount Wilson, CA
13 91023, USA}

\author[0000-0002-1575-4310]{Jacob Ennis}
\affiliation{University of Michigan, Deparment of Astronomy, Ann Arbor, MI 48109, USA}

\author[0000-0001-5980-0246]{Benjamin R. Setterholm}
\affiliation{Astronomy Department, University of Michigan, Ann Arbor, MI 48109, USA}

\author[0000-0002-3003-3183]{Tyler Gardner}
\affiliation{Astronomy Department, University of Michigan, Ann Arbor, MI 48109, USA}

\author[0000-0001-6017-8773]{Stefan Kraus}
\affiliation{Astrophysics Group, Department of Physics \& Astronomy, University of Exeter, Stocker Road, Exeter, EX4 4QL, UK}

\author[0000-0001-9764-2357]{Claire L. Davies}
\affiliation{Astrophysics Group, Department of Physics \& Astronomy, University of Exeter, Stocker Road, Exeter, EX4 4QL, UK}


\author[0000-0001-9647-2886]{Jerome A. Orosz}
\affiliation{San Diego State University, Department of Astronomy, San Diego, CA, 92182 USA}



\begin{abstract}
We present measurements of the interferometrically-resolved binary star system 12 Com and the single giant star 31 Com in the cluster Coma Berenices. 12 Com is a double-lined spectroscopic binary system consisting of a G7 giant and an A3 dwarf at the cluster turnoff. Using an extensive radial velocity dataset and interferometric measurements from PTI and the CHARA array, we measured masses $M_1 = $\pmass \perr $ M_\odot$ and $M_2 = $\smass \serr$ M_\odot$. Interferometry also allows us to resolve the giant, and measure its size as $R_1 = 9.12 \pm 0.12 \pm 0.01 R_\odot$. With the measured masses and radii, we find an age of \primage $ \ \pm \ 41  \pm 42  $ Myr. For comparison, we measure the radius of 31 Com to be \intradthirtyone $R_\odot$. Based on the photometry and radius measurements, 12 Com A is likely the most evolved bright star in the cluster, large enough to be in the red giant phase, but too small to have core helium burning. Simultaneous knowledge of 12 Com A's mass and photometry puts strong constraints on convective core overshooting during the main sequence phase, which in turn reduces systematic uncertainties in the age. Increased precision in measuring this system also improves our knowledge of the progenitor of the cluster white dwarf WD$1216+260$.
\end{abstract}



    \section{Introduction} \label{sec:intro}

Stellar ages are among the most difficult quantities to measure as age is not a directly measurable characteristic, making model interpretation a necessity. Since models are needed to derive an age, then stellar mass becomes an important characteristic to know. When we can determine the mass of an evolved star, age precision often will dramatically increase because these stars have such short evolution timescales. The accuracy of the age will still depend on whether the models have accurate physics in them, and uncertainties in internal parameters such as chemical composition cause additional systematic uncertainties. In this project, we are analyzing evolved stars in binary systems that are themselves within star clusters, allowing for mass determinations of the stars and age determination for the wider cluster.

The Coma Ber open star cluster is the second closest open cluster to the Sun (after Hyades), and its extent is now resolvable in the radial direction with distances measured by {\it Gaia} \citep{gaiaedr3}. 12 Com is a spectroscopic binary member of Coma Ber, composed of a G7 giant primary star and an A3 dwarf main sequence star  with an angular separation of $\sim 20$ mas that can be resolved interferometrically \citep{griffin}. \cite{griffin} presented the most comprehensive study on the orbit of 12 Com. However, masses have not been determined for 12 Com. We seek to use improved radial velocities and interferometry to refine the orbit of the binary.
Characterization of the giant star primary, as evolved as it is, should provide the most precise age to date for the cluster.  

\subsection{Literature Ages}\label{litage}

So far, all age analyses for Coma Ber have involved isochrone analysis of color-magnitude diagrams (CMDs) for the member stars. One of the oldest attempts at dating Coma Ber comes from \cite{tsvetkovage} who used Population I pulsating stars to determine an age of $410 \pm 230$ Myr. \cite{casewellage} used NEXTGEN isochrones \citep{nextgeniso} to determine an age of $450 \pm 50$ Myr. PARSEC isochrones \citep{parsec} were used by \cite{holmberggc} and \cite{singh} to measure ages of $500 \pm 100$ Myr and $700 \pm 100$ Myr, respectively. 

The large differences in age as well as the large associated uncertainties stem from the inherent limitations of isochrone analysis. Isochrone analysis primarily uses the turnoff point and evolved stars in its determination of age. For Coma Ber, the stars around the turnoff point show significant scatter in CMDs similar to that of Hyades and Praesepe \citep{brandt}. As such, it is likely that any analysis solely involving isochrone fitting is underestimating the uncertainties in the age determination. In addition, there are only two giant stars (the giant in 12 Com, and 31 Com) in the whole cluster, and to use 12 Com in an analysis, one must be able to measure the characteristics of the two stars separately. 

\subsection{Distance, Reddening and Metallicity}\label{litdist}

Coma Ber is one of the closest clusters to us, which means there are precise parallax measurements of cluster stars. \cite{bailerjones} determined distances and probability distributions for individual cluster stars from {\it Gaia} Early Data Release 3 \citep{gaiaedr3} parallaxes, including accounting for non-zero parallax zeropoints. The distance to 12 Com was $86.99 \pm 1.19$ pc, which is used below. 

Reddening analysis for the Coma cluster has been conducted by multiple groups \citep[e.g.,][]{singh,taylorreddening} by comparing to objects in Praesepe and Hyades. There is agreement that the foreground reddening of Coma Ber is consistent with zero. \cite{taylorreddening} in particular finds an upper bound for the reddening of Coma Ber of $E(B-V) \le 0.0032$. 

Metallicity measurements of Coma Ber quoted in the literature vary around solar, and analyses have been conducted using different types of stars. \cite{frielmetal} report [Fe/H] $= -0.06 \pm 0.03$ from an analysis of nine F stars, and [Fe/H] $= -0.04 \pm 0.02$ from 5 G-type stars. \cite{gebranmetal} used 14 F-type stars to determine [Fe/H]$ = +0.07 \pm 0.09$. More recently \cite{soutometal} found [Fe/H] $= 0.00 \pm 0.01$, $0.03 \pm 0.01$, $0.04 \pm 0.01$, and $0.04 \pm 0.03$ for 3 F-type, 4 G-type, 4 K-type, and 7 M-type stars, respectively. 

\section{Observations and Data Reduction}

\subsection{Spectroscopy} \label{sec:spec}

\subsubsection{Literature Data}


\cite{griffin} compiled a list of over 300 primary and over 40 secondary star literature radial velocity measurements. The earliest measurements of the primary star belonged to \cite{hansen} who used the Hartmann Spectrocomparator at Lick Observatory. One additional measurement was made at Mt. Wilson using the coud\'{e} spectrograph at the 100 inch telescope \citep{lickrv}; three using the 36-inch refractor at Lick Observatory \citep{lickrv}; 2 by \citet{parsonsrv} using the Coud\'{e} Spectrograph at the 2.1 m Struve telescope at McDonald Observatory; 3 by \citet{beaversrv} using the radial velocity spectrometer at the Fick Observatory; 1 by \citet{cannonrv} using the 15 inch refractor at the Dominion Observatory at Ottawa; and 1 by \citet{Glushkovarv} using a correlation spectrograph at the Zeiss-1000 1m telescope at Mt. Maidanak Observatory. \cite{massarotti} made nine measurements from the Wyeth Reflector at the Oak Ridge Observatory at Harvard. 

\cite{griffin} supplemented the literature measurements with their own observations using the ELODIE spectrograph at Observatoire de Haute-Provence (OHP), the spectrometer at the 200-inch Palomar telescope, as well as the Plaskett 1.83m telescope at the Dominion Astrophysical Observatory \citep{[DAO;][]harper} during almost a 30-year period from 1981 to 2009. The DAO observations in particular also allowed measurements of the secondary component. Griffin also used independent radial velocities obtained from \cite{abtwill} using the coud\'{e} spectrograph on the Kitt Peak 0.9m coud\'{e} telescope. All of the data obtained and compiled by \cite{griffin} were weighted accordingly and added into our analysis below.



\subsubsection{Archival Spectra}

We analyzed 16 additional spectra from the ELODIE archive \citep{moultaka}, originally taken over a 26 month period at the 1.93m telescope at OHP (P.I. X. Delfosse). The spectra have a spectral resolution of $R = 42,000$ with a signal-to-noise (S/N) averaging over 200. The spectra covered $4000 - 6800$ \AA, and were reduced using the ELODIE pipeline. Of these spectra (Spectra 2), one had much lower S/N than the others ($\sim 26$), while another (Spectra 16) had strong atmospheric contamination. Neither was used in our subsequent analysis. The primary star dominates the spectra, and its velocities could be measured with no noticeable effect from the secondary star.

We focused on detection and measurement of the secondary star in the spectra in order to expand its radial velocity dataset and  better constrain the mass of the primary star. While the secondary star is substantially fainter than the primary, its lines can sometimes be distinguished due to a larger rotational broadening ($v_{\rm rot} \sim 200$ km s$^{-1}$) and different spectral type (A3). 

We attempted to separate the spectra of the primary giant and the secondary main sequence star in several ways. We began with spectral disentangling \citep{gonzalez} to split the light contributions from the stars. This involves an iterative determination of average spectra and fitting of broadening functions derived using two different synthetic spectral templates. This did not produce satisfactory results, largely because the luminosity of the primary is so much larger than that of the secondary ($\Delta m > 3$ mag).

Our most successful attempt involved isolating the signal from the secondary star by subtracting a proxy spectrum for the primary star, similar to the method used by \citet{griffin}. We used an ELODIE spectrum of 31 Vul \citep{griffin}, which is very similar to 12 Com A: both are G7 giants around $10-15 R_\odot$ \citep{vul31}. 31 Vul's spectra were scaled and Doppler shifted appropriately before being subtracted from each 12 Com spectra. We used a modified version of {\tt BF-rvplotter}\footnote{https://github.com/mrawls/BF-rvplotter} to calculate broadening functions \citep{rucinskisvd} from the subtracted spectra, and we fit analytic rotational broadening functions to determine the radial velocity of the secondary star in each. This was done by utilizing the ATLAS synthetic spectral template from Pollux \citep{pollux} for the secondary star with a $T_{\rm eff} = 8500$ and log $g = 3.5$.

\begin{figure}[h]
	\centering
	\includegraphics[width=\linewidth]{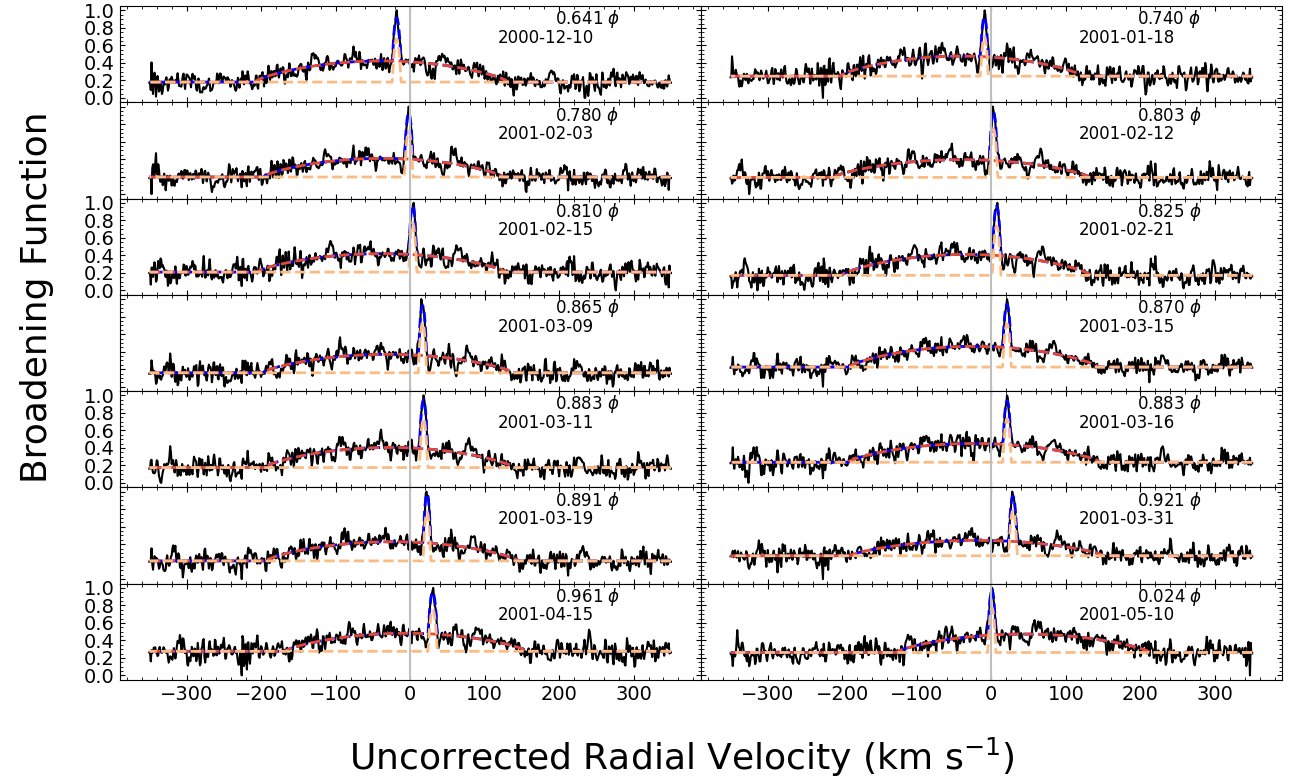}
	\caption{Broadening functions of 14 ELODIE spectra after the primary star's signal was subtracted from the composite spectra. The broadening functions are normalized to the peak value. The blue curve is the fit to the remnant unsubtracted signal from the primary star, while the secondary star's fit is shown in red. Observation dates and associated phase are shown in each panel.
	}
	\label{broadfunc}
\end{figure}

We examined two subsets of spectral lines for the fitting of the secondary spectra --- the Balmer lines and the Mg II lines around 4481 \AA. Thanks to the earlier spectral type of the secondary star, the Balmer lines clearly stand out in the raw spectra, and have high signal-to-noise ratio. Mg lines were primarily used in velocity measurements by \cite{griffin}, and although they have a much smaller signal-to-noise ratio than the Balmer lines, they are less affected by pressure broadening. We find that velocities measured when including the Balmer lines were systematically offset toward the system velocity compared to those measured with Mg lines alone. This may be related to the additional pressure broadening involved in the Balmer lines. In our analysis below, we use velocities measured in the $4460 - 4560$ \AA \hspace{1pt} spectral range, surrounding the Mg II lines. For the fourteen usable spectra, the broadening functions and fits are shown in Figure \ref{broadfunc}. Both the residual signal from the primary and the signal from the secondary can be seen. The secondary star's lines are much more broadened than the primary's, corresponding to an average rotational speed of 175 km s$^{-1}$. 
The ELODIE radial velocities are given in Table \ref{tab:radvel} and are heliocentrically corrected.

\begin{deluxetable*}{ccc}
\tablecaption{New Radial Velocity Measures for 12 Com  \label{tab:radvel}}
\tablewidth{0pt}
\tablehead{
\colhead{mJD\tablenotemark{a}} &\colhead{$v_A$ (km/s)} &\colhead{$v_B$ (km/s)}  
}
\startdata
51888.68786  &  $8.90\pm0.21$  &  $-13.66\pm2.27$  \\  
51927.66765  &  $14.68\pm0.21$  & $-19.92\pm2.32$  \\  
51943.69036  &  $16.06\pm0.21$  & $-21.36\pm1.41$  \\  
51952.68066  &  $17.41\pm0.22$  & $-28.66\pm2.63$  \\  
51955.64859  &  $17.74\pm0.21$  &  $-28.12\pm2.01$  \\  
51961.61818  &  $18.61\pm0.21$  &  $-26.01\pm2.20$  \\  
51977.58259  &  $20.82\pm0.21$  &  $-29.73\pm2.54$  \\  
51979.51683  &  $21.04\pm0.21$  &  $-24.02\pm2.26$  \\  
51984.48075  &  $21.52\pm0.22$  &  $-30.70\pm2.35$  \\  
51984.52174  &  $21.52\pm0.22$  &  $-25.84\pm2.51$  \\  
51987.53708  &  $21.62\pm0.21$  &  $-28.35\pm2.44$  \\  
51999.50986  &  $22.26\pm0.21$  &  $-28.78\pm1.26$  \\  
52015.49332  &  $16.06\pm0.22$  &  $-21.81\pm2.61$  \\  
52040.43123  &  $-20.40\pm0.21$  &  $18.22\pm2.59$  \\  
\enddata
\tablenotetext{a}{mJD = BJD - 2400000} 
\end{deluxetable*}

\textbf{\subsection{Data Acquistion}}

\begin{figure}[h]
\includegraphics[width=0.9\linewidth]{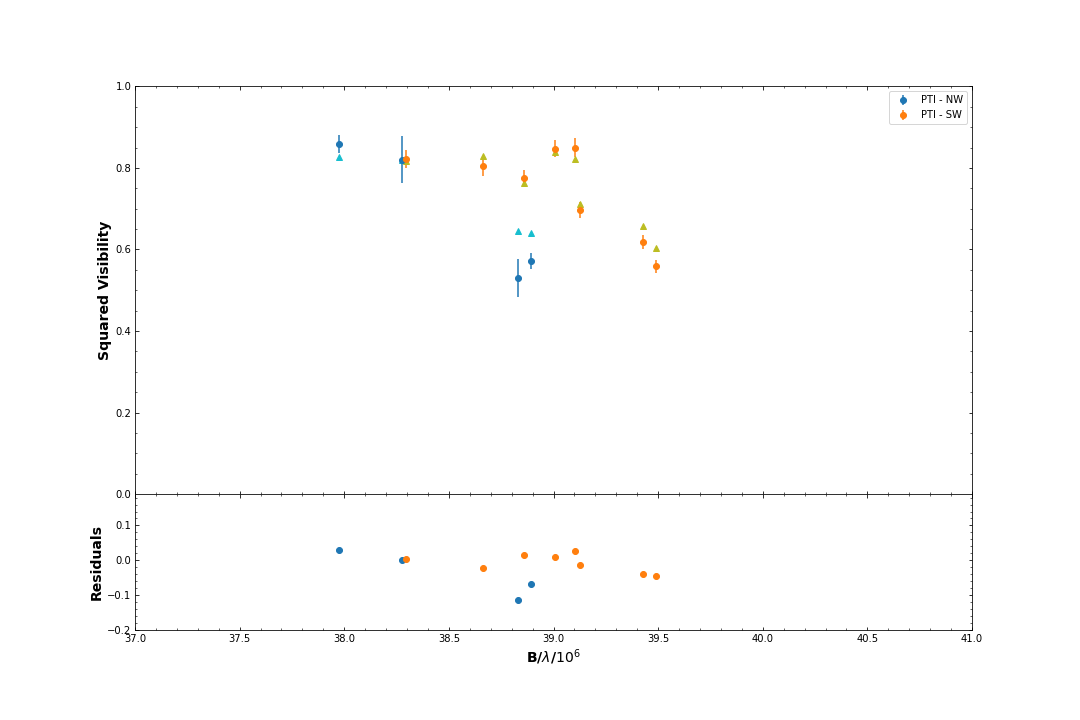}
\caption{Visibility data and corresponding model fit for the PTI data on 12 Com. All data were taken in a two-week period. Circles and triangles refer to the data and model, respectively. Colors refer to different telescope pairs. The bottom plot shows the residuals between the model and data.}
\label{visfitpti}
\end{figure}

We obtained a small number of archival observations taken over the course of two weeks in 2004 using the Palomar Testbed Interferometer \citep[PTI;][]{pti}. The PTI observations were extracted from the NASA Exoplanet Science Institute (NExScI) archive. The observations used two-telescope configurations, and so only produced visibility measurements. The wideband $K$ observations were calibrated against two other stars observed on the same nights, and the visibilities were computed using the software {\tt wbCalib}\footnote{https://nexsci.caltech.edu/software/V2calib/wbCalib/index.html}. These observations, shown in Figure \ref{visfitpti}, provide our only constraint on  the eastern end of the astrometric orbit (see Figure \ref{skyplot}).

A larger number of observations were obtained using the \mircx beam combiner \citep{mircx} at the Center for High Angular Resolution Astronomy (CHARA) Array \citep{chara} at the Mt. Wilson Observatory. The \mircx instrument interferes the light from up to six telescopes, resulting in visibility measurements on 15 baselines and 20 closure phase triangles. All \mircx measurements were taken in the $H$ band using the $R = 50$ (Prism 50), $R = 102$ (Prism 102), or $ R = 190$ (Grism 190) spectral modes. Visibilities and closure phases were obtained using the standard \mircx pipeline version 1.3.3 - 1.3.5\footnote{https://gitlab.chara.gsu.edu/lebouquj/mircx\_pipeline}. The log for the PTI and \mircx measurements is given in Table \ref{interfdata}, and the measurements of visibilities and closure phases are given in Tables \ref{tab:vis} and \ref{tab:phi}.

We also obtained \mircx observations of the other cluster giant (31 Com) for comparison of radii with 12 Com A. We obtained observations on four nights in March and April 2022, outlined in Table \ref{interfdata}. Although we measured visibilities and closure phases, all of the closure phase measurements were consistent with zero, as expected for a point-symmetric object.

\subsection{Interferometry}

We briefly summarize the observable quantities we measured and fit.
The visibility amplitude $|\mathcal{V}|$ is defined by how clearly the fringe pattern can be observed. In the case of a uniform disk with angular diameter $\alpha$, this is:
\begin{equation}\label{bessel}
   | \mathcal{V} | = \frac{2J_1(\pi B \alpha /\lambda)}{\pi B \alpha/\lambda}
\end{equation}
$J_1$ is the Bessel function of the first kind, and $B$ is the projected baseline of the interferometer based on the star's position in the sky. A binary system can be approximated by two uniform disks, and the observable (squared visibility) is given by:  
\begin{equation}\label{binviseq}
	\mathcal{V}^2_{bin} = \frac{\mathcal{V}_P^2 + r^2 \mathcal{V}_S^2 + 2 r |\mathcal{V}_P||\mathcal{V}_S| \cos [2\pi \vec{B} \cdot \vec{s}/\lambda]}{(1+r)^2}
\end{equation}
Here $\mathcal{V}_P $ and $\mathcal{V}_S$ refer to the complex visibilities of the primary and secondary stars, respectively. The luminosity ratio in the observed wavelength band is given by $r$. $\vec{B}$ is the baseline vector for the interferometer, and $\vec{s}$ is the angular separation vector of the two stars in the sky. 

\begin{figure}[h]
\includegraphics[width=0.9\linewidth]{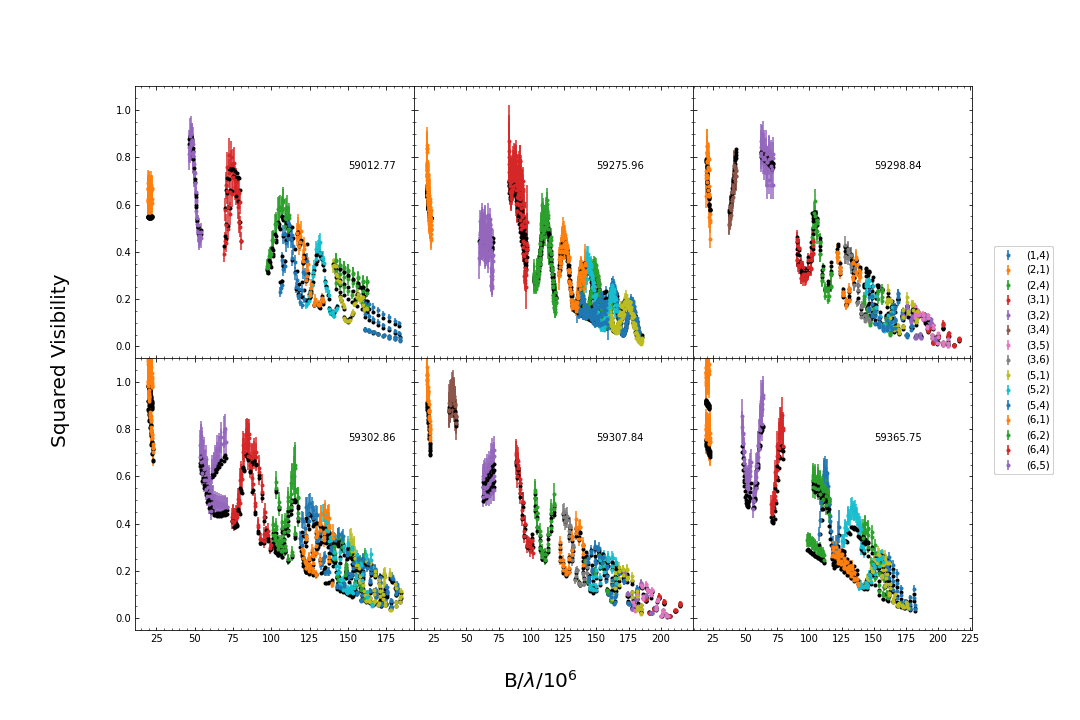}
\caption{Visibility data and corresponding model fit for the \mircx data on 12 Com. The visibilities are split by observation date, where the dates are HJD$ -2400000 $. Colors refer to different telescope pairs.}
\label{visfit}
\end{figure}

Fringe visibilities are a function of the projected baseline for the interferometer. By utilizing multiple visibilities taken at the same epoch, we can get information about the separation, orientation, and flux ratio of two stars in a binary system. As can be seen for the \mircx data in Figure \ref{visfit}, several signatures are present in each epoch. The oscillations as a function of spatial frequency reveal information about the angular separation of the stars and their orientation on the sky. The amplitude from maximum to minimum of the visibility oscillation gives the binary flux ratio. In addition, the clear downward trend with spatial frequency is an illustration that the angular diameter of one of the stars is resolved. It is worth noting, however, that systematic uncertainties in the wavelengths of the observations affect the determination of the slope of the visibility curve, which introduce a systematic uncertainty when converting the spatial frequencies to angular scales.

\begin{figure}[h]
\includegraphics[width=0.9\linewidth]{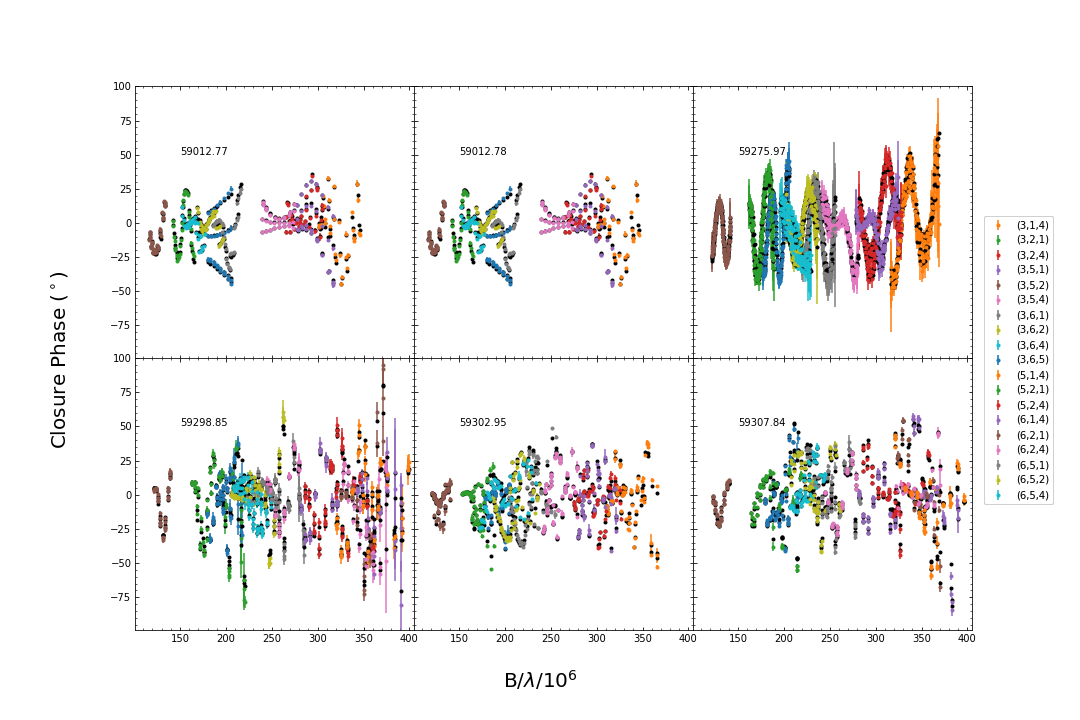}
\caption{Closure phase data and corresponding model fit for the \mircx data on 12 Com. The closure phases are split by observation date, as seen in each panel, where the dates are HJD$ -2400000 $. Colors refer to different triplets of telescopes.}
\label{phifit}
\end{figure}

While visibilities can be affected by atmospheric turbulence, we can calculate another observable that is invariable to atmospheric perturbations, namely the triple product (or more specifically the argument of the triple product) known as the closure phase \citep{monnier}. This quantity measures asymmetry in the light distribution and is well suited to determine the luminosity ratio in any particular band of stars in a binary system. A single point source would have a closure phase of 0$^\circ$. The closure phases of 12 Com measured with the \mircx beam combiner are shown in Figure \ref{phifit}. The maximum and minimum deviations of the closure phases from zero provides the luminosity ratio in the associated wavelength band. As with the visibility amplitudes, the periodicity of the closure phases gives the binary separation and orientation.

\begin{deluxetable}{c cccccc l  }
\tablecaption{Log of Interferometric Observations }
\tablehead{
\colhead{Facility} & \colhead{Combiner} & \colhead{HJD Start} & \colhead{UT Date}  & \colhead{Spectral Mode} & \colhead{$N_{vis}$} & \colhead{$N_{cp}$} & \colhead{Calibrators}}
\startdata
\multicolumn{8}{c}{12 Com} \\
Palomar &  PTI (SW) & 2453127 & 5/02/04 & & 2 &  & HD 108471, 108722 \\
Palomar &  PTI (SW) & 2453132 & 5/07/04 & & 3 &  & HD 108471, 108722 \\
Palomar &  PTI (SW) & 2453139 & 5/14/04 & & 3 &  & HD 108471, 108722 \\
Palomar &  PTI (NW) & 2453140 & 5/15/04 & & 2 &  & HD 108471, 108722 \\
Palomar &  PTI (NW) & 2453142 & 5/17/04 & & 2 &  & HD 108471, 108722 \\
CHARA & \mircx &  2459012 &  6/12/20  & Prism 50 & 320 & 320 & HD 108382\\
CHARA & \mircx &  2459275 &  3/02/21  & Grism 190 & 1398 & 1400 & HD 107966, 108382\\
CHARA & \mircx &  2459298 &  3/25/21  & Prism 50 & 480 & 640 & HD 108382\\
CHARA & \mircx &  2459302 &  3/29/21  & Prism 50 & 640 & 560 & HD 107966, 108382, 113771 \\
CHARA & \mircx &  2459307 &  4/03/21  & Prism 50 & 480 & 640 & HD 108382, 113771\\
CHARA & \mircx &  2459365 &  5/31/21  & Prism 50 & 464 & 396 & HD 108382, 113771\\
\hline
\multicolumn{8}{c}{31 Com} \\
CHARA & \mircx & 2459655 & 3/18/22 & Prism 50 & 481 & 640 & HD 113771\\
CHARA & \mircx & 2459675 & 4/06/22 & Prism 50 + Prism 102 & 841 & 1120 & HD 107655, 113771, 116233\\
CHARA & \mircx & 2459677 & 4/08/22 & Prism 50 & 401 & 480 & HD 113771\\
CHARA & \mircx & 2459678 & 4/09/22 & Prism 102 & 961 & 1280 & HD 113771, 116233\\
\enddata
\tablecomments{Assumed uniform disk calibrator angular diameters: HD 108471: 0.64 mas; HD 108722: 0.48 mas; HD 106661: 0.341 mas; HD 107655: 0.187 mas; HD 107966: 0.358 mas; HD 108382: 0.401 mas; 
HD 113771: 0.420 mas; 
HD 116233: 0.188 mas \citep{jmmc}}
\label{interfdata}
\end{deluxetable}

\begin{deluxetable}{c|cccc|cc|c  }[h]
\tablecaption{Visibilities for 12 Com  \label{tab:vis}}
\tablewidth{0pt}
\tablehead{
\colhead{mJD\tablenotemark{a}} &\colhead{$\lambda (\mu m)$ } &\colhead{$\sigma_\lambda$} &\colhead{$\mathcal{V}^2$} &\colhead{ $\sigma_{\mathcal{V}}$}  
&\colhead{$u$ (m)} &\colhead{$v$ (m)} &\colhead{Configuration} }
\startdata
53127.71391 & 2.2110 & 0.4000 & 0.8799 & 0.0128 &  -40.71931 &   76.02423 & SW \\ 
53127.76668 & 2.2104 & 0.4000 & 0.5775 & 0.0153 &  -53.25317 &   69.15584 & SW \\ 
53132.69574 & 2.2264 & 0.4000 & 0.8416 & 0.0191 &  -39.41604 &   76.52089 & SW \\ 
53132.76592 & 2.2276 & 0.4000 & 0.7214 & 0.0157 &  -55.46800 &   67.23171 & SW \\ 
53132.81003 & 2.2232 & 0.4000 & 0.8504 & 0.0203 &  -60.21185 &   60.18388 & SW \\ 
53139.67251 & 2.2110 & 0.4000 & 0.8055 & 0.0195 &  -38.20224 &   76.95870 & SW \\ 
53139.72202 & 2.2126 & 0.4000 & 0.6394 & 0.0182 &  -50.89806 &   70.85630 & SW \\ 
53139.77027 & 2.2100 & 0.4000 & 0.8838 & 0.0248 &  -58.55339 &   63.55237 & SW \\ 
53140.66747 & 2.2198 & 0.4000 & 0.5846 & 0.0237 &  -83.51411 &  -21.87902 & NW \\ 
53140.72152 & 2.2162 & 0.4000 & 0.8951 & 0.0227 &  -77.03038 &  -33.90183 & NW \\ 
53142.65695 & 2.2190 & 0.4000 & 0.5209 & 0.0642 &  -83.63297 &  -20.71664 & NW \\ 
53142.70889 & 2.2168 & 0.4000 & 0.8407 & 0.0919 &  -78.42691 &  -32.37164 & NW \\ 
59012.76881 & 1.7083 & 0.0301 & 0.9002 & 0.0757 &   77.73756 &   25.37662 & E2-S1 \\ 
59012.76881 & 1.6819 & 0.0301 & 0.8721 & 0.0734 &   77.73756 &   25.37662 & E2-S1 \\ 
59012.76881 & 1.6516 & 0.0301 & 0.8191 & 0.0691 &   77.73756 &   25.37662 & E2-S1 \\ 
59012.76881 & 1.6206 & 0.0301 & 0.7347 & 0.0621 &   77.73756 &   25.37662 & E2-S1 \\ 
59012.76881 & 1.5890 & 0.0301 & 0.6104 & 0.0519 &   77.73756 &   25.37662 & E2-S1 \\ 
59012.76881 & 1.5568 & 0.0301 & 0.5121 & 0.0438 &   77.73756 &   25.37662 & E2-S1 \\ 
59012.76881 & 1.5240 & 0.0301 & 0.4663 & 0.0400 &   77.73756 &   25.37662 & E2-S1 \\ 
59012.76881 & 1.4983 & 0.0301 & 0.4827 & 0.0414 &   77.73756 &   25.37662 & E2-S1 \\ 
59012.76881 & 1.7083 & 0.0301 & 0.3440 & 0.0300 & -112.74443 &  125.09320 & E2-W2 \\ 
\enddata
\tablecomments{Table \ref{tab:vis} is published in its entirety in the machine-readable format. A portion is shown here for guidance regarding its form and content.}
\tablenotetext{a}{mJD = HJD - 2400000}
\end{deluxetable}

\begin{deluxetable}{c|cccccccc|c  }[h]
\tablecaption{Closure Phases for 12 Com  \label{tab:phi}}
\tablewidth{0pt}
\tablehead{
\colhead{mJD\tablenotemark{a}} &\colhead{$\lambda$ ($\mu$m) } &\colhead{$\sigma_{\lambda}$} &\colhead{$\phi$ (${}^\circ$) } &\colhead{$\sigma_\phi$}  &\colhead{$u_1$(m)}&\colhead{$v_1$(m)}&\colhead{$u_2$(m)}&\colhead{$v_2$ (m)} &\colhead{Configuration}  }
\startdata
59012.76881 & 1.7083 &  0.0301 &  -4.34 &   1.09 &   77.73756 &  25.3766 & -190.48198 &   99.71658 & S2-S1-E2 \\ 
59012.76881 & 1.6819 &  0.0301 &  -7.72 &   0.86 &   77.73756 &  25.3766 & -190.48198 &   99.71658 & S2-S1-E2 \\ 
59012.76881 & 1.6516 &  0.0301 &  -4.65 &   0.86 &   77.73756 &  25.3766 & -190.48198 &   99.71658 & S2-S1-E2 \\ 
59012.76881 & 1.6206 &  0.0301 &   1.80 &   0.86 &   77.73756 &  25.3766 & -190.48198 &   99.71658 & S2-S1-E2 \\ 
59012.76881 & 1.5890 &  0.0301 &   8.47 &   0.86 &   77.73756 &  25.3766 & -190.48198 &   99.71658 & S2-S1-E2 \\ 
59012.76881 & 1.5568 &  0.0301 &   8.42 &   0.86 &   77.73756 &  25.3766 & -190.48198 &   99.71658 & S2-S1-E2 \\ 
59012.76881 & 1.5240 &  0.0301 &  -2.06 &   0.86 &   77.73756 &  25.3766 & -190.48198 &   99.71658 & S2-S1-E2 \\ 
59012.76881 & 1.4983 &  0.0301 & -11.68 &   1.27 &   77.73756 &  25.3766 & -190.48198 &   99.71658 & S2-S1-E2 \\ 
59012.76881 & 1.7083 &  0.0301 &   1.40 &   0.86 &   77.73756 &  25.3766 & -208.69421 &  127.74149 & S2-S1-E1 \\ 
59012.76881 & 1.6819 &  0.0301 &   2.45 &   0.86 &   77.73756 &  25.3766 & -208.69421 &  127.74149 & S2-S1-E1 \\ 
59012.76881 & 1.6516 &  0.0301 &  -1.21 &   0.86 &   77.73756 &  25.3766 & -208.69421 &  127.74149 & S2-S1-E1 \\ 
59012.76881 & 1.6206 &  0.0301 & -18.99 &   0.86 &   77.73756 &  25.3766 & -208.69421 &  127.74149 & S2-S1-E1 \\ 
59012.76881 & 1.5890 &  0.0301 & -33.66 &   0.92 &   77.73756 &  25.3766 & -208.69421 &  127.74149 & S2-S1-E1 \\ 
59012.76881 & 1.5568 &  0.0301 & -21.27 &   0.86 &   77.73756 &  25.3766 & -208.69421 &  127.74149 & S2-S1-E1 \\ 
59012.76881 & 1.5240 &  0.0301 &   4.12 &   1.24 &   77.73756 &  25.3766 & -208.69421 &  127.74149 & S2-S1-E1 \\ 
59012.76881 & 1.4983 &  0.0301 &  24.77 &   2.05 &   77.73756 &  25.3766 & -208.69421 &  127.74149 & S2-S1-E1 \\ 
59012.76881 & 1.7083 &  0.0301 &  -2.19 &   0.86 &   77.73756 &  25.3766 & -125.86196 & -135.89354 & S2-S1-W2 \\ 
\enddata
\tablecomments{Table \ref{tab:phi} is published in its entirety in the machine-readable format. A portion is shown here for guidance regarding its form and content.}
\tablenotetext{a}{mJD = HJD - 2400000}
\end{deluxetable}


\subsection{Spectral Energy Distributions (SEDs)}

\begin{figure}[h]
\includegraphics[width=0.9\linewidth]{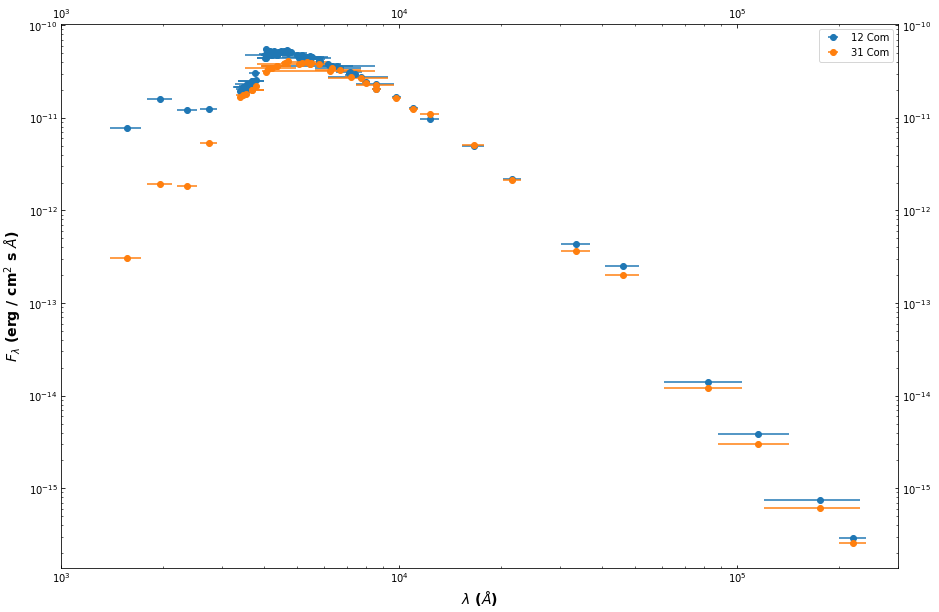}
\caption{Spectral energy distributions of the single giant star 31 Com and the binary 12 Com, with the photometric passbands for each point indicated with horizontal bars.
}
\label{sedcomp}
\end{figure}

While not the primary focus of this paper, spectral energy distributions were obtained for 12 Com and the other Coma Ber giant star 31 Com from published photometry. More discussion of the data can be found in Appendix \ref{app:sed}.
When we compare the two SEDs (see Figure \ref{sedcomp}), the binary's SED shows much greater ultraviolet emission than 31 Com. The indications from the analysis below are that 12 Com A is cooler than 31 Com, which emphasizes that this ultraviolet contribution must be coming from a relatively hot companion star.

\section{Interferometry and Radial Velocity Modeling} \label{sec:orbit}


\subsection{12 Com}

Before conducting a simultaneous fit to the full interferometric and spectroscopic dataset, we first used the Eclipsing Light Curve \citep[ELC;][]{elc} code to fit for the systemic radial velocities of the two stars in different spectroscopic datasets. We wish to ensure
that the radial velocities are on the same velocity zeropoint, as this has some effect on fitted orbital velocities and stellar masses. \cite{griffin} reported that \textbf{they} artificially increased OHP and ESO velocity measurements by 0.8 km s$^{-1}$ in order to keep the Cambridge zero-point velocities \citep{cambridgezp} for all of the datasets he used. We find that we need to apply an offset of $+0.52$ km s$^{-1}$ to the ELODIE dataset for the primary star to match the \citeauthor{griffin} observations, and this is fairly consistent with the Cambridge velocity zero-point. The comparison of secondary star datasets indicates that we need to apply an offset of $-6.43$ km s$^{-1}$, however.  The secondary star offset is fairly large, and may reflect some residual difficulties in measuring the secondary star in the presence of the large primary star signal or measurement differences due to different lines used.


We model the binary's orbit using \texttt{interfRVorbit}, which uses a genetic algorithm \citep{chargenetic} to search for the best binary orbital fit to interferometric measurements (visibilities and closure phases), radial velocities for both stars, and sky position data. Fourteen parameters were fitted: radial velocity semi-amplitudes of each component ($K_p, K_s$), eccentricity ($e$), argument of periastron ($\omega$), orbital period ($P$), reference time for periastron ($t_P$), luminosity ratios in the $H$ and $K_S$ bands ($r_H, r_K$), angular diameter of the primary star ($\alpha_P$), orbit inclination ($i$), angular size of the semi-major axis of the binary orbit ($a$), position angle of the ascending node ($\Omega$), and systemic velocities for both stars ($\gamma_p, \gamma_s$). While we clearly resolve the primary in the interferometric data, the secondary star is not resolved. The angular size of the secondary does have a small effect on the fits, however, and we force the size of the secondary star to be 0.315 mas, based on visible-light interferometric measurements of other Coma Ber stars with similar luminosities and temperatures \citep{perraut}. We initially fitted for the $H$- and $K$-band luminosity ratios independently, and the best fits resulted in values that differed by about 25\%. However, because the interferometric measurements were taken in wavelength bands well onto the Rayleigh-Jeans portion of the spectral energy distribution, the luminosity ratios are expected to effectively be equal. For that reason, we enforced the equality of the ratios in our fits. Analysis of the interferometric observations indicates that the luminosity ratio $L_2 / L_1$ in $H$ and $K$ infrared bands is approximately 0.14. Finally, we allowed for different systemic velocities in the fit due to the difference in evolutionary states between the two stars. As discussed in Section \ref{sec:ages}, one star is a giant, and one is an A-type main sequence star. Gravitational redshifts (larger for the main sequence star) and convective blueshifts (probably only important for the giant) can produce differences of order one to a few km s$^{-1}$.

Because different types of observational data can comment on the values of some parameters, it is important to properly weight the distinct datasets. We attempted to do this by empirically determining scalings for the uncertainties that came out of the data reduction process. As much as possible, we fitted subsets of the observed data separately, and scaled the uncertainties to return a reduced $\chi_\nu^2 \approx 1$. Effectively, this uses observed scatter around the best-fit model as an empirical measure of the uncertainty. For example, we fitted radial velocity datasets from different sources separately to independently determine the scalings for those data. 

Overall, 10000 genetic generations were made to assess uncertainties for each parameter by exploring ranges around the best fit values. The orbital parameters implied by this fit are given in Table \ref{parmresults}, as are the fitting results of \cite{griffin} to their radial velocity dataset. We obtained uncertainty measurements for the fit parameters by examining parameter ranges at one above the minimum value of $\chi^2$ mark \citep{avnichi}.

\begin{figure}[h]
\centering
\includegraphics[width=\linewidth]{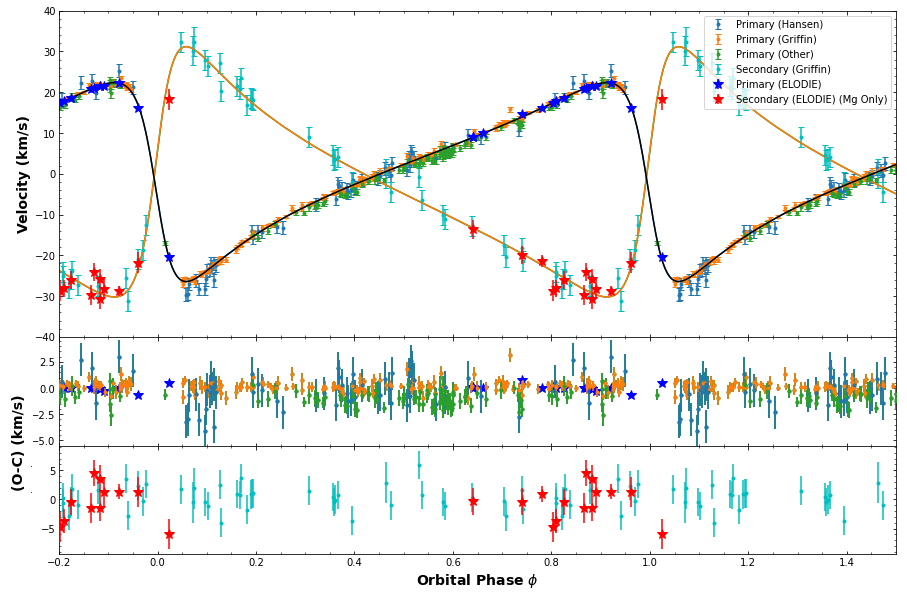}
\caption{Top: Radial velocity measurements from ELODIE (red and blue stars) and literature data versus phase, along with the best-fit orbit model. Middle/bottom: Velocity residuals (observed minus computed) for the primary and secondary stars, respectively. Primary star measurements from \cite{griffin} between orbital phases of $-0.05$ and 0.05 were removed due to likely contamination by the secondary star. }
\label{rvcurve}
\end{figure}

Figure \ref{rvcurve} compares the various radial velocity datasets to the fit. With the new set of observations from ELODIE, particularly in the secondary star's radial velocities, we obtain an improved constraint for the velocity amplitude of the secondary star, and thus the mass of the primary. 

\begin{deluxetable}{ ccccccc   }
\tablecaption{Fitted Positions for 12 Com} 
\tablehead{\colhead{UT Date} &  mJD   &  $\rho$  &  $\theta$    &  $\sigma_{\rm maj}$ & $\sigma_{\rm min}$ & $\varphi$\\
& & (mas) & (deg) & (mas) & (mas) & (deg)}
\startdata
6/12/2020 & 59012.273 & 17.427 & 73.55 & 0.087 & 0.035 & 56.57   \\
3/02/2021 & 59275.465 & 14.140 & 350.91 & 0.071 & 0.041 & 67.56\\
3/25/2021 & 59298.344 & 13.562 & 6.50 & 0.068 & 0.046 & 107.48\\
3/29/2021 & 59302.367 & 13.524 & 9.36 & 0.068 & 0.042 & 41.01\\
3/29/2021 & 59302.384 & 13.521 & 9.34 & 0.068 & 0.052 & 58.40\\
3/29/2021 & 59302.445 & 13.526 & 9.46 & 0.068 & 0.035 & 56.71\\
4/03/2021 & 59307.343 & 13.452 & 12.81 & 0.067 & 0.047 & 111.92\\
5/31/2021 & 59365.258 & 14.877 & 51.74 & 0.075 & 0.011 & 37.21\\
5/31/2021 & 59365.303 & 14.877 & 51.81 & 0.074 & 0.031 & 46.65\\
\enddata
\tablecomments{Column 2: mJD = HJD - 2400000. Column 3: angular separation. Column 4: position angle of separation vector measured East of North. Columns 5-7: the major and minor axis size, and orientation angle of the error ellipse.}
\label{gailpositions}
\end{deluxetable}

\begin{figure}[h]
\centering
\includegraphics[width=.8\linewidth]{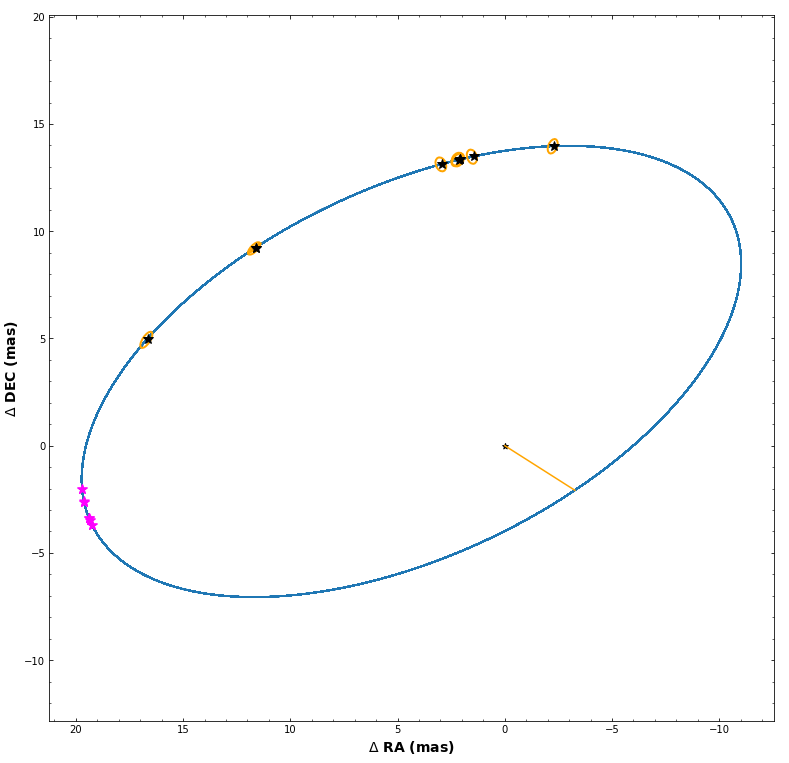}
\caption{Relative orbit for 12 Com B with respect to 12 Com A, represented by the star at (0,0). Purple and black points shows model-predicted positions of the secondary star at the time of PTI and CHARA observations, respectively. Positions derived from the visibilities are coincident with the model-predicted positions. Orange ellipses show a 10$\sigma$ uncertainty around the calculated positions. The yellow line connects 12 Com A to the periastron.}
\label{skyplot}
\end{figure}

The best fitting model of the binary system's astrometric orbit is shown in Figure \ref{skyplot}. The positional measurements obtained from CHARA are outlined in Table \ref{gailpositions} and were computed using a binary grid search procedure \citep{schaefer} \footnote{https://www.chara.gsu.edu/analysis-software/binary-grid-search}. These position measurements are determined from visibility and closure phase measurements on multiple baselines on single nights, are plotted in black. These positions have associated uncertainties that account for up to 0.06\% in the \mircx wavelength scales \citep{gardner}. This systematic error source can be estimated to produce an uncertainty in distance of 0.05 pc. In the same plot we show the expected position of the secondary star at the time of the PTI observations to give an indication of how these help constrain the orbit size. Because we are constraining the orbit with measurements in two quadrants of the orbit with different instruments, we should be aware of the possibility that the wavelength calibrations might lead to systematics in the orbit determination, but we are not presently able to more to evaluate them more than this. 


\begin{deluxetable}{| c |c| c|   }
\tablecaption{Fitted Parameters for 12 Com} 
\tablehead{\colhead{Parameter}   & \colhead{This Study} & \colhead{\cite{griffin}}}
\startdata
$ K_P $ (km s$^{-1}$) & $ 24.399 \pm 0.036 $ & $ 24.40 \pm 0.06 $ \\
$ K_S $ (km s$^{-1}$) &  $ 30.7 \pm 0.4 $ & $ 30.6 \pm 0.4 $ \\
$ P $ (days) &$  396.4473 \pm 0.0002  $& $ 396.411 \pm 0.009 $ \\
$ e $ &   $ 0.599483 \pm 0.000026 $ & $ 0.5978 \pm 0.014 $ \\
$ i $ ($ ^\circ $) &$ 64.8556 \pm 0.0011 $ & \\
$ \omega_1$  ($ ^\circ $) \tablenotemark{a}& $ 100.162 \pm 0.001 $ & $ 100.22 \pm 0.27 $ \\
$ t_P $ (mJD\tablenotemark{b}) &$ 46877.148 \pm 0.054 $ & $ 46877.57 \pm 0.17 $ \\
$ r_H = \frac{L_S}{L_P} (H)$  & $ 0.14423 \pm 0.00020 $ & \\
$ r_K = \frac{L_S}{L_P} (K)$  & $ 0.14423 \pm 0.00020 $ & \\
$ \alpha_p $ (mas) & $ 0.97145 \pm 0.00046 $\tablenotemark{c} & \\
$ a $(mas) & $ 20.3358 \pm 0.00066 $ \tablenotemark{c} &  \\ 
$ \Omega $ ($ ^\circ $ ) & $  118.618 \pm 0.004  $ & \\
$\gamma_P$ (km s$^{-1}$) &  $ 0.51 \pm 0.02  $ & $ 0.78 \pm 0.04 $ \\
$\gamma_S$ (km s$^{-1}$) & $ -2.75 \pm 0.20 $ & \\
\hline
d (pc) & $86.78 \pm 0.76 $ & \\ 
$M_P\sin^3 i$ (M$_\odot$) & $1.958 \pm 0.054$ & $1.97 \pm 0.07 $ \\
$M_S\sin^3 i$ (M$_\odot$) & $1.556 \pm 0.023$ & $1.567 \pm 0.027 $ \\
$ M_P $ ($ M_\odot $) & $ 2.64 \pm 0.07 $ & \\
$ M_S $ ($ M_\odot $) & $  2.10 \pm 0.03 $ & \\
\enddata
\tablenotetext{a}{ Based on the longitude of periastron defined from the spectroscopically measured ascending node}
\tablenotetext{b}{mJD = HJD - 2400000}
\tablenotetext{c}{Uncertainty purely statistical and does not take into account the 0.06\% in the \mircx wavelength scale}
\label{parmresults}
\end{deluxetable}

We note that the systemic velocity of the secondary star appears to be significantly blueshifted ($\sim 3$ km s$^{-1}$) relative to that of the primary, and that this is in the opposite direction from expectations based on differences in gravitational redshift or convective blueshift for the two stars. Although we do not have an explanation for this at present, what is most important for the majority of our analysis here are the velocity amplitudes, as these are what enter into the mass determinations.

\subsection{31 Com}

For the single giant 31 Com, we fitted the interferometric visibilities with a model disk of uniform brightness using two parameters: the angular diameter of the star ($\alpha$) and the normalization of the visibilities ($\mathcal{V}^2_0$). The second parameter was found to be necessary because the calibrated visibilities consistently reached values above 1 on the smallest baselines. We experimented to check whether our calibration targets (listed in Table \ref{interfdata}) may have had incorrect angular diameters, but did not find an explanation for this. 

\begin{figure}[h]
\centering
\includegraphics[width=.8\linewidth]{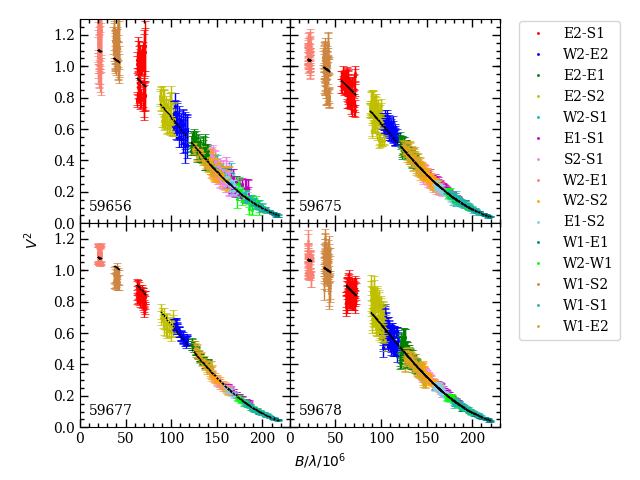}
\caption{Fits to the interferometric data for the giant 31 Com. Color points show observations with different baselines in the CHARA array, and black points show best model fits. Labels give the Julian date of observation - 2400000.}
\label{31comvis}
\end{figure}

We fitted data from each of four nights of observation separately using the \texttt{interfRVorbit} code, as shown in Figure \ref{31comvis}. There was good consistency between the angular diameters from the nights, and the weighted average value is $\alpha = 0.922 \pm 0.004\pm0.005$ mas as shown in Table \ref{31comsizes}. The quoted uncertainties are statistical (based on the standard deviation of the 4 measurements) and systematic (assuming 0.06\% uncertainty in the \mircx wavelength scale). Limb darkening was not included; however, were it to be included, it would likely result in a larger radius measurement and associated uncertainty. A slightly larger radius would not affect our main conclusions.

\begin{deluxetable}{ccc}
\tablecaption{Fitted Parameters for 31 Com} 
\tablehead{\colhead{mJD\tablenotemark{a}}   & \colhead{Visibility Normalization Factor} & \colhead{Angular size (mas)}}
\startdata
59656 &   $1.125 \pm 0.007$     &  $0.9195\pm0.0017$   \\
59675 &   $1.064\pm0.005$      &  $0.9218\pm0.0012$ \\
59677 &   $1.104 \pm 0.005$     &  $0.9274\pm0.0011$ \\
59678 &   $1.090 \pm0.003$       &  $0.9174\pm0.0010$ \\
\enddata
\tablenotetext{a}{mJD = HJD - 2400000}
\label{31comsizes}
\end{deluxetable}

\section{Discussion} \label{sec:ages}

While model isochrone analyses have been done many times before for Coma Ber (see section \ref{litage}), we can now apply new stellar data to the problem. The masses and radius derived from the fit for 12 Com can put strict limits on the age of the cluster. This also provides insight into the fidelity of the internal physics used in different isochrone models, and can identify issues that will allow us to minimize systematic errors. In particular, convective core overshooting is notoriously difficult to model, with different modeling groups utilizing different amounts of overshooting. This directly affects the amount of hydrogen burned on the main sequence, and thus the age.

Data on the brightest stars in the Coma Ber cluster are very helpful to use in connection with the binary star information from 12 Com. We created an initial list of probable members from a {\it Gaia} EDR3 sample. Stars with $G < 18$ were considered candidates if positions were within $15^\circ$ of $(\alpha, \delta) = (186\fdg38, 25\fdg42)$, proper motions were within 2.5 mas yr$^{-1}$ of $(\mu_\alpha, \mu_\delta) = (-12.1,-9.0)$ mas yr$^{-1}$, and parallaxes were within 0.3 mas of 11.64 mas \citep{gaiaedr3}. We also incorporated stars in the cluster's tidal tails from the more thorough surveys by \citet{tang} and \citet{furnkranz}. When available, we also used radial velocity information from \citet{merm} or {\it Gaia} DR2 \citep{gaiadr2} to check membership (using the mean $v_r = 0.21 \pm0.13$ km s$^{-1}$; \citeauthor{gaiaedr3}).

The cluster contains only two bright stars that have evolved off the main sequence --- one of the stars of 12 Com, and 31 Com. 31 Com is a single rapidly-rotating variable subgiant in the Hertzsprung gap. \citet{strassmeier} identified major starspots (revealing a rotational period of 6.8 d) and a large rotational velocity ($v_{rot} \sin i = 67 \pm2$ km s$^{-1}$).
We can also derive the effective temperature $T_{\rm eff} \approx 5700 \pm 40$ K by comparing the SED of 31 Com with synthetic spectra (see Figure \ref{seds} and Appendix \ref{app:sed}).
The bolometric flux can also be derived from the SED fit combined with the {\it Gaia} distance in order to get a luminosity of \lumthirtyone $L_\odot$. 
Pairing this with the effective temperature, we find that 31 Com has a radius of \lumradthirtyone $R_\odot$. 
This measurement is in rough agreement with the radius derived from the interferometry (\intradthirtyone$ R_\odot$). Typically, a large scatter around the long baseline end implies asymmetric structures or stellar spots. As such, the lack of scatter in the visibilities of 31 Com in Figure \ref{31comvis} suggests that 31 Com is nearly circular in shape. These data show that 31 Com is clearly smaller in size than 12 Com A, but despite the rapid evolution timescale for these giants, the radii are fairly similar. If an asteroseismic determination of its mass could be accomplished, the star could also provide a strong constraint on the cluster age.

\subsection{Mass-Radius Isochrones}

One of the results of this paper is that interferometric observations have resolved the primary star 12 Com A. Using the angular diameter of the primary with the Gaia distance for 12 Com ($86.99 \pm 1.19$ pc; \citealt{bailerjones}), we find a primary star radius of $9.12 \pm 0.12 \pm 0.01 R_\odot$, where the first and second uncertainties are estimates of the statistical and systematic uncertainties, respectively. The statistical uncertainty includes the contributions from the model fit for the angular diameter as well as the Gaia distance. The systematic uncertainty derives from a 0.06\% uncertainty in the \mircx interferometer wavelength scale. 

\begin{figure}[h]
	\centering
	\includegraphics[width=0.9\linewidth]{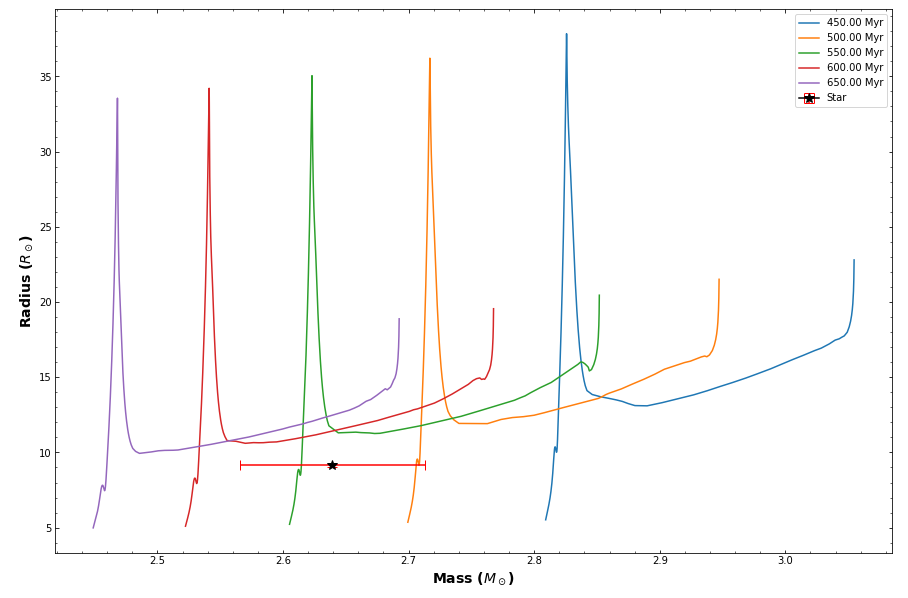}
	\caption{Mass-radius comparison of 12 Com A with MIST isochrones \citep{mist1} showing only the sub-giant and red giant branch phases. }
	\label{mriso}
\end{figure}

This radius tells us that even though the binary's color is bluer than 31 Com, the primary star in the binary is larger than 31 Com, and thus is a clue that it is likely to be redder and slightly more evolved.
We compare 12 Com A's characteristics with MIST isochrones \citep{mist1} in the mass-radius plane in Figure \ref{mriso}. With the relatively large uncertainty for the mass, the radius constrains the age of the cluster to about $538\pm38$ Myr. The radius measurement does rule out the possibility that the primary star is in the red clump phase, which would require it to be significantly larger, and in fact, constrains the evolutionary phase to near the luminosity minimum at the base of the red giant branch (where there is a small kink in the isochrones). 

\subsection{Evolutionary States of the Stars}

\begin{figure}[h]
	\centering
	\includegraphics[width=0.9\linewidth]{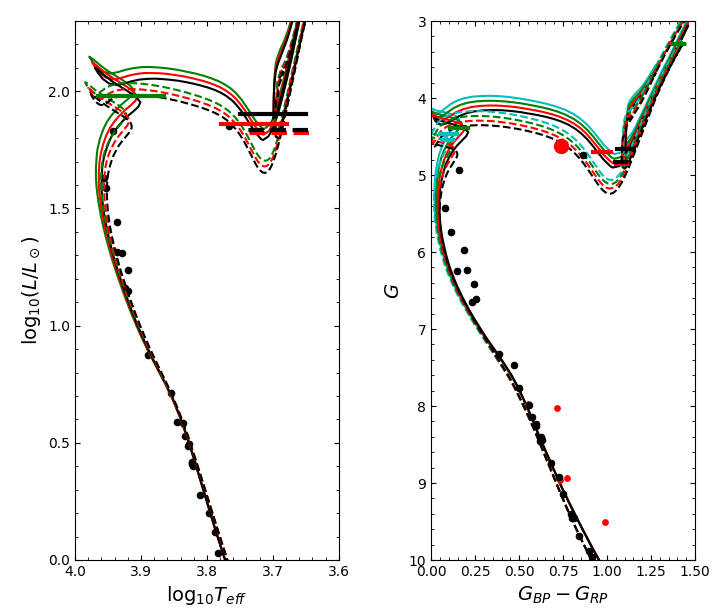}
	\caption{MIST (dashed lines; \citealt{mist1}) and PARSEC v. 1.2 (solid lines; \citealt{parsec}) isochrones with $Z = 0.0152$ and $ E(B-V) = 0$.  The horizontal lines are the model positions of the primary star (assuming $2.64 M_\odot$) in each isochrone. 
	{\it Left:} HR Diagram for Coma Ber stars. The isochrones have ages of 520 (green), 540 (red), and 560 (black) Myr. 
	{\it Right:} A Gaia color-magnitude diagram for Coma Ber stars. The isochrones have ages of $500 - 560$ Myr in 20 Myr increments. The large top-most red point shows the combined light of} 12 Com. Smaller red points show known binary systems.
	
	\label{isotypecomp}
\end{figure}

Figure \ref{isotypecomp} shows theoretical isochrones from the PARSEC \citep{parsec} and MIST \citep{mist1} databases (shown with solid and dotted lines, respectively) plotted against data for cluster members. This comparison is done with PARSEC v1.2 rather than newly introduced 2.0. The latter version uses a smaller maximum core overshooting parameter ($\lambda_{ov,max} = 0.4$ rather than $\lambda_{ov,max} = 0.5$), as well as a diffusive treatment for convective mixing. This results in a hotter and slightly fainter MS phase and a fainter subgiant phase \citep{parsec2.0}. The comparison of isochrones emphasizes that uncertainty in the amount of convective core overshooting affects the measured age of the cluster by tens of Myr.

\begin{figure}[h]
	\centering
	\includegraphics[width=0.9\linewidth]{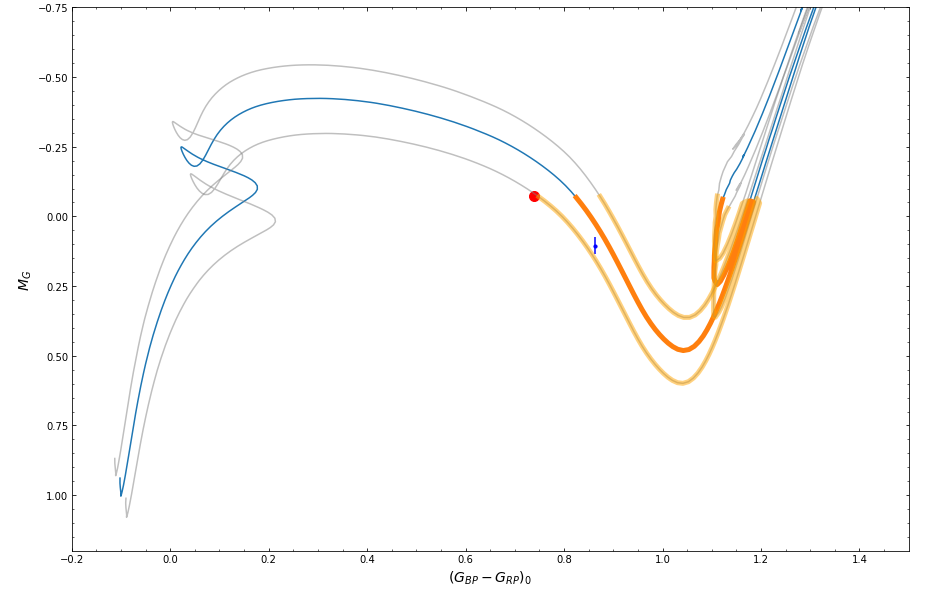}
	\caption{Evolutionary tracks (MESA) for a \pmass M$_\odot$ star (blue curve) with [Fe/H] $ = 0.07$.  The combined photometry for 12 Com is shown with a red dot, using a distance of $87.0 \pm 0.2$ pc and reddening $E(B-V) = 0.00$ $(E(BP-RP) = 0.0022)$. The photometric position of 31 Com is shown with a  blue point. The dark orange curve shows the allowed regions for the giant to exist based on photometric considerations (ages $ 531-535 $ Myr for giant branch, $ 554 - 634 $ Myr for red clump). Evolution tracks for masses at $\pm 1\sigma$ (\perr $M_\odot$) are shown with grey/light orange curves.}
	\label{etpossrange}
\end{figure}

As 12 Com is a binary system, the photometry of the stars in the binary must be disentangled for comparisons with evolutionary models. The two components of 12 Com obviously must be dimmer than the binary, and one component must be redder than the binary's overall color. This on its own constrains the evolutionary state of the more evolved star to be a late subgiant, an early red giant, or a red clump star, as shown in Figure \ref{etpossrange}. With our measurement of the radius above, we can rule out the possibility of it being in the red clump.

\begin{figure}[h]
	\centering
	\includegraphics[width=0.9\linewidth]{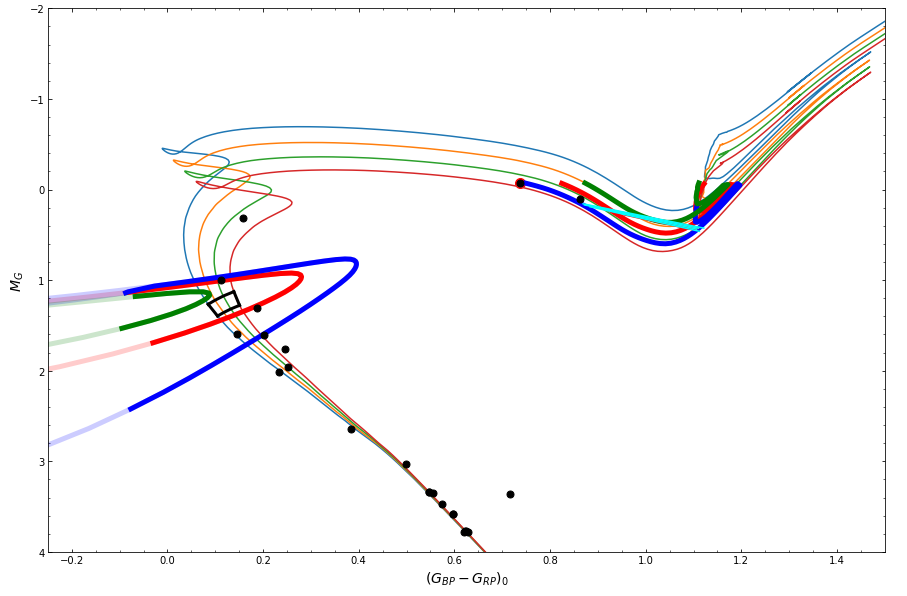}
	\caption{Gaia color-magnitude diagram for Coma Ber stars (black points), assuming reddening $E(B-V) = 0$ and distance modulus $(m-M)_0 = 4.67$. MIST isochrones \citep{mist1} are shown for 450 Myr (blue) to 600 Myr (red) with 50 Myr spacing.  Thick curves on the right show allowed portions of evolution tracks (fainter and redder than the binary) for the measured mass of 12 Com A as well as $\pm1 \sigma$  away in red, green, and blue, respectively. Corresponding positions for the secondary star (when 12 Com A's photometry is subtracted from that of the binary) are shown on the left. The cyan box shows the required position of 12 Com A if the secondary is in the isochrones between the thick blue lines. The black box encloses the predicted positions of the secondary in the isochrones, given its mass and $\pm1 \sigma$ uncertainties. 
	}
	\label{secphoto}
\end{figure}

The age uncertainty is dominated by the uncertainty in the primary star mass because the evolution timescale for the radius is so short in the giant phase. We can further constrain the age by examining the photometry implied for the secondary. By subtracting the allowed photometry for the primary from that of the binary, the secondary star can be localized to a small region near the cluster turnoff point, as shown in Figure \ref{secphoto}. The light red curve shows a large range of potential colors and magnitudes, but only those close to the main sequence should be considered if the star has evolved like a normal single star cluster member. This in turn puts somewhat tighter limits on where 12 Com A can be on its evolution track and also slightly reduces the age range allowed. The predicted positions of 12 Com B in the isochrones for its mass (and $\pm1 \sigma$ uncertainty, shown by the black box) do agree with the limits placed by this procedure. 

By assuming that 12 Com B is in the intersection between the MIST isochrones and the thick blue curve in Figure \ref{secphoto} (imposed by the allowed photometric positions of 12 Com A for $M_1 - 1\sigma$), we can further delimit the color-magnitude area that 12 Com A can exist in. This range is shown as the cyan box, and it constrains the age to \primage \primageerr \ Myr.

To fit the {\it Gaia} CMD of the brightest main sequence stars in the cluster, we find some need to use MIST or PARSEC isochrones with super-solar metal content. The issue is complicated, however, due to continuing uncertainty as to the precise value of the solar metal content $Z_\odot$. MIST isochrones and evolutionary tracks use a protosolar metal content $Z = 0.0142$, consistent with the solar composition study of \citet{asplund}. A recent study by \citet{magg} find that solar constraints are much better matched by $Z_\odot = 0.0177$. For the purposes of reproducibility, we will quote the best fit value used in the isochrones: $Z = 0.0165$. If $Z_\odot$ is the higher \citeauthor{magg} value, then Coma Ber would appear to have subsolar metal content ([Fe/H] $= -0.03$). If the lower \citeauthor{asplund} is correct, then Coma Ber would appear to have super-solar metallicity ([Fe/H] $= +0.07$). In either case, the metallicity [Fe/H] would be within the range covered by spectroscopic abundance determinations for the cluster in the literature (see section \ref{litdist}).

Literature values for metallicity of Coma stars tend to vary by about $\pm 0.08$ dex. Utilizing different evolutionary tracks given this metallicity range results in an age difference about $\pm14$ Myr. 
Combining the uncertainties due to mass and metallicity in quadrature, we find that the age of the evolved star 12 Com A is constrained to \primage $ \pm 41$ Myr.

\subsection{Convective Core Overshooting}

One the major theoretical uncertainties in stellar evolution models is the treatment of convective core overshooting. Models from different groups utilize different overshoot lengths (parameterized in units of the pressure scale height), but they are generally assumed to ramp up from zero to a plateau value as a function of stellar mass \citep{claret}. Stars with the masses of the stars in 12 Com are assumed to have overshoot of the plateau value. We have used the PARSEC v1.2 and v2.0 isochrones to estimate the effects of uncertainties in convective core overshooting on the cluster age. Taking the overshoot in v1.2 to be approximately 0.25 pressure scale heights ($H_P$) in version 1.2 and approximately 0.2 scale heights in v2.0 \citep{parsec2.0}, we can see the effects on the luminosity of the subgiant branch that is inhabited by 12 Com A and 31 Com. PARSEC v1.2 and v2.0 isochrones differ by as much as 70 Myr for the CMD position of 12 Com A, but the $0.05 H_P$ difference in convective core overshooting distances probably does not represent the uncertainty properly.

\cite{claret} semi-empirically found from eclipsing binary stars that the overshoot length for stars with $M > 2 M_\odot$ was $0.20 H_P$, with an uncertainty of $0.03 H_P$ for evolved stars (similar to 12 Com A). Taking this to be the best representation of the core overshooting uncertainty, we derive an associated systematic age uncertainty of 42 Myr. 

\subsection{White Dwarf Initial-Final Mass Relation} 

An area of study that benefits immensely from an improved age is the white dwarf initial-final mass relation (IFMR). Coma Ber is known to have a massive white dwarf member (WD $1216+260$) with a measured "final" mass of $0.90 \pm 0.04 M_\odot$ \citep{donnie}. A precise age can be used to derive the ``initial" mass of the star before it became a white dwarf. \cite{donnie} used a cluster age of $500 \pm 100$ Myr to infer a progenitor star mass of $4.77^{+5.37}_{-0.97} M_\odot$. The large uncertainty is due primarily to uncertainty in the cluster age, as derived from the CMD analysis. As our understanding of the giant star mass loss prior to white dwarf emergence is poor, the precision measurement of a star's mass shortly before its major final mass loss is important \citep{kaliwhitedwarf}.

With our new age measurement, we can greatly reduce the uncertainty in the initial mass. Using the \citeauthor{donnie} cooling age $\tau_{cool} = 363^{+46}_{-41}$ Myr and our new cluster age, we get an inferred progenitor star mass of $4.35^{+1.94}_{-0.73} M_\odot$, where the uncertainty includes statistical and systematic (core overshooting) contributions. This cleanly rules out the possibility that the progenitor star had a mass near the white dwarf/neutron star production boundary. This gives us a much better picture of the progenitor star right before major mass loss.

\section{Conclusions} \label{sec:conclusion}

Ages and masses are always useful in astronomy, and the age for an open cluster like Coma Ber can serve as a benchmark. 
Almost all age measurements for Coma Ber thus far have involved the analysis of color-magnitude diagrams \citep{leeuwenage,casewellage,tsvetkovage,strassmeierage,singh}, with significant uncertainties that result from a lack of constraints on some of the physics that affects the evolution of the stars.
This paper marks the first look into the cluster age using an analysis of one of the most evolved stars in the cluster, thereby providing an important check on CMD analyses .
Mass and age measurements for the evolved stars in 12 Com could not be made previously because the orbital inclination had not been determined. We take advantage of CHARA and PTI interferometric data to fit 12 Com's astrometric orbit and measure the masses of the primary and secondary stars: \pmass \perr $M_\odot$ and \smass \serr $M_\odot$, respectively. Using the mass of the primary with MIST evolutionary tracks, we determine the cluster age using the photometry we derived for the primary and secondary stars. With our interferometric measurements, we were also able to resolve the primary star and measure the radius as $9.12 \pm 0.12 R_\odot$. These results restrict the age of the cluster to \primage$ \pm 41 \pm 42$ Myr. Major contributions to the statistical uncertainty come from the mass uncertainty (due to a small number of secondary star radial velocities and large measurement uncertainties), and to a lesser extent, from metallicity uncertainty.  We have also examined the effects of overshooting and how much it contributes to a systematic uncertainty in the age.




Reliably-measured ages are necessary for other applications, such as honing the initial-final mass relation for white dwarfs and determining the absolute calibration of rotation period as an age indicator for isolated field main sequence stars \citep{barnes}. Such gyrochronology studies generally use color as a stand-in for mass, and have utilized standard isochrone fitting for ages. With significant improvements in precision for ages as well as reliable measurements for mass, the empirical models can be refined. The rotational properties of Coma Ber stars are not as well studied as for clusters like the Hyades, but ground-based measurements have been presented in \citet{collier} and \citet{terrien}, and a combination of ground-based and {\it TESS} measurements were presented in \citet{singh}. 
Even without the absolute age calibration, gyrochronological studies are capable of relative age measurements, with \citeauthor{collier} finding that Coma was consistent with being the same age as the Hyades (Coma Ber $34\pm41$ Myr younger), and \citeauthor{singh} finding that Coma was coeval with the Hyades and Praesepe, despite differences in chemical composition between the clusters.

\software{BF-rvplotter (https://github.com/mrawls/BF-rvplotter)}

\begin{acknowledgments}
We gratefully acknowledge support from the National Science Foundation under grant AAG 1817217 to E.L.S.
This research made use of observations from the SIMBAD database, operated at CDS, Strasbourg, France; and the WEBDA database, operated at the Institute for Astronomy of the University of Vienna. 

This work is based upon observations obtained with the Georgia State University Center for High Angular Resolution Astronomy Array at Mount Wilson Observatory. The CHARA Array is supported by the National Science Foundation under Grant No. AST-1636624 and AST-2034336.  Institutional support has been provided from the GSU College of Arts and Sciences and the GSU Office of the Vice President for Research and Economic Development. Time at the CHARA Array was granted through the NOIRLab community access program (NOAO PropID: 2020A-0379; NOIRLab PropIDs: 2021A-0276, 2022A-118161; PI: E. Sandquist). \mircx received funding from the European Research Council (ERC) under the European Union's Horizon 2020 research and innovation programme (Starting Grant "ImagePlanetFormDiscs" No. 639889). SK acknowledges support from ERC Consolidator Grant "GAIA-BIFROST" (Grant No. 101003096) and STFC Consolidator Grand ST/V000721/1. JDM acknowledges funding for the development of \mircx (NASA-XRP NNX16AD43G, NSF-AST 1909165) and MYSTIC (NSF-ATI 1506540, NSF-AST 1909165).
The Palomar Testbed Interferometer was operated by the NASA Exoplanet Science Institute and the PTI collaboration. It was developed by the Jet Propulsion Laboratory, California Institute of Technology with funding provided from the National Aeronautics and Space Administration. This work has made use of services produced by the NASA Exoplanet Science Institute at the California Institute of Technology. This research has made use of the Jean-Marie Mariotti Center Aspro and SearchCal services.
\end{acknowledgments}

%

\facilities{PO:PTI, CHARA (\mircx), OHP:1.93m (ELODIE)}

\clearpage


\appendix
\section{SEDs of 12 Com and 31 Com}\label{app:sed}


We utilize a variety of sources for the SEDs of 31 Com. In the ultraviolet part of the spectrum, we obtained photometry from the Sky Survey Telescope on the TD-1 satellite \citep{td1uv}, which consisted of four pass bands (centered at 1565 \AA, 1965 \AA, 2365 \AA, and 2740 \AA). 

In the optical portion of the spectrum, we obtained Str\"{o}mgren \textit{uvby} photometry from \cite{paunzen}. These magnitudes were then calibrated using reference fluxes from \cite{gray}. We obtained photometry from the Johnson-13 color system as described in \cite{johnsondata} and the Tycho filters $B_T$ and $V_T$ from the Tycho Reference catalogue \citep{tycho2}. Both were calibrated using reference fluxes taken from the Spanish Virtual Observatory (SVO) Filter Profile Service \citep{rodrigo}. We also obtained photometry in the Johnson-Cousins filters \citep{moreldata}, which were calibrated to flux using Table A2 of \cite{besselldata}. Four photometric measurements in the $WBVR$ 4-color system were taken by \cite{wbvrphoto} and calibrated with zeropoints from \cite{mann}. Finally, we used the high precision {\it Gaia} data in the $G, G_{BP}$ and $G_{RP}$ bandpasses from Early Data Release 3 \citep{gaiaedr3}.

For the infrared part of the spectrum, we utilized photometry in the $JHK_s$ bands from the Two-Micron All-Sky Survey \citep[2MASS;][]{2mass}. We utilized the reference fluxes from \cite{cohen} to convert to fluxes. Photometry from four bands ($W1, W2, W3$, and $W4$) in the Wide Field Infrared Survey Explorer \citep[WISE;][]{wise} were taken and converted using their own reference fluxes. Finally, in the infrared range, we obtained fluxes from the AKARI satellite IRC all-sky survey in the $S9W$ and $L18W$ filters \citep{akari}. 

A comprehensive list of all the photometric measurements used is shown in Table \ref{tab:photometry}.
All the above photometric sources are also utilized for 12 Com. However, 12 Com has a few additional sources, mostly in the optical. In particular, we obtained observations from the Geneva Observatory \citep{geneva}, the seven-filter Vilnius photometry \citep{vilnius}, spectrophotometric measurements from \cite{clampitt}, and SDSS \citep{sdss}.

\begin{figure}[h]
	\centering
	\includegraphics[width=0.9\linewidth]{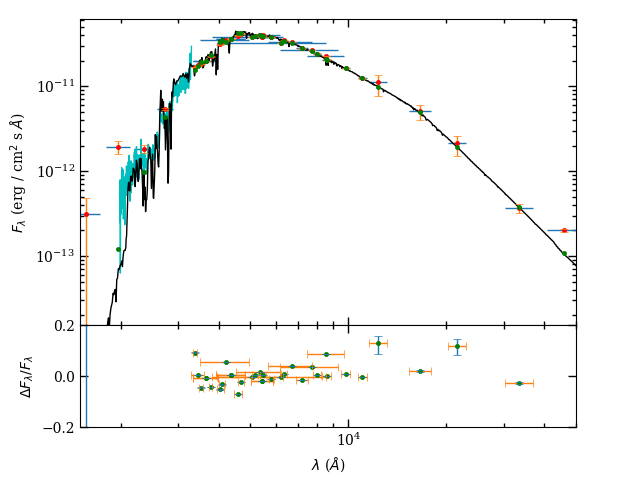}
	\caption{{\it Top panel:} Spectral energy distribution of the single giant star 31 Com (orange points), with the photometric passbands for each point indicated with horizontal bars. The best fit ATLAS9 model with $T_{\rm eff} = 5700$ K and $\log g = 3.5$ is shown in black, and the model predictions for each observed bandpass are shown with green points. A flux-calibrated IUE spectrum is shown in cyan. {\it Bottom panel:} Relative flux residuals for each observed photometric bandpass.
    }
    \label{seds}
\end{figure}

To get properties of 31 Com, we fit the photometric SED using ATLAS9 \citep{atlas9} models with [Fe/H] $= 0.0$ and $\log g = 3.5$. This model fit is shown in Figure \ref{seds}. This resulted in an effective temperature of approximately 5700 K and a bolometric flux $F_{\rm bol} = 2.86 \times 10^{-7}$ erg cm$^{-2}$ s$^{-1}$. We estimate statistical uncertainties of $\pm40$ K in $T_{\rm eff}$ from scatter in infrared-flux method temperatures from $J$, $H$, and $K_s$ bands, and of $\pm3 \times 10^{-9}$ erg cm$^{-2}$ s$^{-1}$ in $F_{\rm bol}$ using fits with different $T_{\rm eff}$ within the range. Metallicity uncertainties contribute negligibly in these fits. 31 Com is known to be magnetically active and an X-ray and UV emitter, so we do not worry greatly about the disagreement between photometric observations and models for $\lambda < 3000$ \AA. An IUE spectrum of the star (data I.D.: LWR04860; P.I.: R. F. Garrison) shows approximate agreement with the fitted model through the ultraviolet. 
Pairing the bolometric flux with a distance of 86.99 pc corresponds to a luminosity $L = $ \lumthirtyone $ L_\odot$, typical of red giant branch stars. This luminosity then corresponds to a radius of \lumradthirtyone $R_\odot$. This agrees very well with the radius derived from interferometry of \intradthirtyone $ R_\odot$.

\clearpage 

\startlongtable
\begin{deluxetable*}{cc|ccc|ccc|c  }
\tablecaption{Photometry of 31 Com and 12 Com  \label{tab:photometry}}
\tablewidth{0pt}
\tablehead{
\colhead{} & \colhead{} & \colhead{} & \colhead{31 Com} & \colhead{} &\colhead{} &\colhead{12 Com} &\colhead{} &\colhead{}  \\
\hline
\colhead{Filter} & \colhead{$\lambda_{eff}$ (\AA)} & \colhead{$m_{\lambda}$} & \colhead{$\sigma_{m}$} & \colhead{$F_\lambda$} & \colhead{$m_{\lambda}$} & \colhead{$\sigma_{m}$} & \colhead{$F_\lambda$} & Notes \\
\colhead{} & \colhead{(\AA)} & & & \colhead{(erg / cm$^2$ s \AA)} & & & \colhead{(erg / cm$^2$ s \AA)} & 
}
\startdata
$F1565$ & 1565 & 10.10 & 0.841 & $ 3.10 \times 10^{-13} $ & 6.61 & 0.048 & $ 7.68 \times 10^{-12} $ & 1  \\ 
$F1965$ & 1965 & 8.11 & 0.186 & $ 1.93 \times 10^{-12} $ & 5.81 & 0.036 & $ 1.61 \times 10^{-11} $ & 1  \\ 
$F2365$ & 2365 & 8.16 & 0.106 & $ 1.84 \times 10^{-12} $ & 6.10 & 0.034 & $ 1.22 \times 10^{-11} $ & 1  \\ 
$F2740$ & 2740 & 7.00 & 0.022 & $ 5.38 \times 10^{-12} $ & 6.09 & 0.015 & $ 1.24 \times 10^{-11} $ & 1  \\ 
$j33$ & 3374 & 5.74 & 0.010 & $ 1.68 \times 10^{-11} $ & 5.59 & 0.010 & $ 1.94 \times 10^{-11} $ & 2  \\ 
$u$ & 3447 & 7.06 & 0.011 & $ 1.75 \times 10^{-11} $ & 6.89 & 0.007 & $ 2.05 \times 10^{-11} $ & 3  \\ 
$U_V$ & 3450 & & & &  7.35 &  & $ 2.15 \times 10^{-11} $ & 11  \\ 
$U_G$ & 3471 & & & &  6.08 &  & $ 2.13 \times 10^{-11} $ & 12  \\ 
$m3500$ & 3500 & & & &  6.61 & 0.016 & $ 2.07 \times 10^{-11} $ & 13  \\ 
$j35$ & 3537 & 5.59 & 0.010 & $ 1.82 \times 10^{-11} $ & 5.51 & 0.010 & $ 1.97 \times 10^{-11} $ & 2  \\ 
$u_{SDSS}$ & 3551 & & & &  6.47 & 0.060 & $ 2.31 \times 10^{-11} $ & 14  \\ 
$W_4$ & 3554 & & & &  5.51 &  & $ 2.20 \times 10^{-11} $ & 15  \\ 
$m3571$ & 3571 & & & &  6.59 & 0.016 & $ 2.04 \times 10^{-11} $ & 13  \\ 
$m3636$ & 3636 & & & &  6.43 & 0.016 & $ 2.27 \times 10^{-11} $ & 13  \\ 
$U$ & 3663 & 5.81 & 0.004 & $ 1.98 \times 10^{-11} $ & 5.56 & 0.009 & $ 2.50 \times 10^{-11} $ & 7  \\ 
$U$ & 3663 & 5.81 &  & $ 1.98 \times 10^{-11} $ & 5.57 &  & $ 2.47 \times 10^{-11} $ & 4  \\ 
$P_V$ & 3740 & & & &  6.69 &  & $ 3.02 \times 10^{-11} $ & 11  \\ 
$j37$ & 3774 & 5.71 & 0.010 & $ 2.23 \times 10^{-11} $ & 5.56 & 0.010 & $ 2.58 \times 10^{-11} $ & 2  \\ 
$B1_G$ & 4023 & & & &  5.46 &  & $ 4.43 \times 10^{-11} $ & 12  \\ 
$m4036$ & 4036 & & & &  5.22 & 0.016 & $ 5.59 \times 10^{-11} $ & 13  \\ 
$j40$ & 4046 & 5.89 & 0.010 & $ 3.12 \times 10^{-11} $ & 5.51 & 0.010 & $ 4.43 \times 10^{-11} $ & 2  \\ 
$X_V$ & 4054 & & & &  5.91 &  & $ 4.88 \times 10^{-11} $ & 11  \\ 
$v$ & 4100 & 6.02 & 0.009 & $ 3.39 \times 10^{-11} $ & 5.62 & 0.005 & $ 4.91 \times 10^{-11} $ & 3  \\ 
$m4167$ & 4167 & & & &  5.22 & 0.016 & $ 5.26 \times 10^{-11} $ & 13  \\ 
$B_T$ & 4220 & 5.72 & 0.014 & $ 3.49 \times 10^{-11} $ & 5.38 & 0.014 & $ 4.77 \times 10^{-11} $ & 5  \\ 
$B_G$ & 4246 & & & &  4.43 &  & $ 4.87 \times 10^{-11} $ & 12  \\ 
$m4255$ & 4255 & & & &  5.19 & 0.016 & $ 5.21 \times 10^{-11} $ & 13  \\ 
$B$ & 4361 & 5.61 &  0.005 & $ 3.60 \times 10^{-11} $ & 5.29 & 0.005 & $ 4.82 \times 10^{-11} $ & 7  \\ 
$B$ & 4361 & 5.61 & & $ 3.60 \times 10^{-11} $ & 5.30 &  & $ 4.79 \times 10^{-11} $ & 4  \\ 
$B_4$ & 4382 & & & &  5.30 &  & $ 5.03 \times 10^{-11} $ & 15  \\ 
$m4400$ & 4400 & & & &  5.12 & 0.016 & $ 5.20 \times 10^{-11} $ & 13  \\ 
$B2_G$ & 4482 & & & &  5.78 &  & $ 5.27 \times 10^{-11} $ & 12  \\ 
$m4565$ & 4565 & & & &  5.01 & 0.016 & $ 5.31 \times 10^{-11} $ & 13  \\ 
$j45$ & 4586 & 5.46 & 0.010 & $ 3.92 \times 10^{-11} $ & 5.18 & 0.010 & $ 5.08 \times 10^{-11} $ & 2  \\ 
$Y_V$ & 4665 & & & &  5.24 &  & $ 5.44 \times 10^{-11} $ & 11  \\ 
$g_{SDSS}$ & 4686 & & & &  4.98 & 0.030 & $ 5.04 \times 10^{-11} $ & 14  \\ 
$b$ & 4688 & 5.38 & 0.019 & $ 4.15 \times 10^{-11} $ & 5.12 & 0.004 & $ 5.30 \times 10^{-11} $ & 3  \\ 
$m4785$ & 4785 & & & &  4.96 & 0.016 & $ 5.09 \times 10^{-11} $ & 13  \\ 
$m5000$ & 5000 & & & &  4.93 & 0.016 & $ 4.77 \times 10^{-11} $ & 13  \\ 
$G_{BP}$ & 5051 & 5.09 & 0.003 & $ 3.80 \times 10^{-11} $ & 4.92 & 0.003 & $ 4.45 \times 10^{-11} $ & 6  \\ 
$Z_V$ & 5162 & & & &  4.95 &  & $ 4.80 \times 10^{-11} $ & 11  \\ 
$j52$ & 5180 & 5.09 & 0.010 & $ 3.90 \times 10^{-11} $ & 4.94 & 0.010 & $ 4.49 \times 10^{-11} $ & 2  \\ 
$m5263$ & 5263 & & & &  4.88 & 0.016 & $ 4.52 \times 10^{-11} $ & 13  \\ 
$V_T$ & 5350 & 5.00 & 0.009 & $ 4.03 \times 10^{-11} $ & 4.86 & 0.009 & $ 4.59 \times 10^{-11} $ & 5  \\ 
$V1_G$ & 5402 & & & &  5.53 &  & $ 4.58 \times 10^{-11} $ & 12  \\ 
$V_V$ & 5442 & & & &  4.76 &  & $ 4.71 \times 10^{-11} $ & 11  \\ 
$V$ & 5448 & 4.94 & 0.007 & $ 3.84 \times 10^{-11} $ & 4.80 & 0.018 & $ 4.35 \times 10^{-11} $ & 7  \\ 
$V$ & 5448 & 4.94 &  & $ 3.84 \times 10^{-11} $ & 4.81 &  & $ 4.32 \times 10^{-11} $ & 4  \\ 
$y$ & 5480 & 4.94 & 0.026 & $ 3.95 \times 10^{-11} $ & 4.79 & 0.006 & $ 4.51 \times 10^{-11} $ & 3  \\ 
$V_G$ & 5504 & & & &  4.78 & 0.028 & $ 4.56 \times 10^{-11} $ & 12  \\ 
$V_4$ & 5519 & & & &  4.80 &  & $ 4.50 \times 10^{-11} $ & 15  \\ 
$j58$ & 5806 & 4.76 & 0.010 & $ 3.79 \times 10^{-11} $ & 4.67 & 0.010 & $ 4.09 \times 10^{-11} $ & 2  \\ 
$G_G$ & 5814 & & & &  5.86 &  & $ 4.31 \times 10^{-11} $ & 12  \\ 
$m5840$ & 5840 & & & &  4.76 & 0.016 & $ 4.11 \times 10^{-11} $ & 13  \\ 
$r_{SDSS}$ & 6166 & & & &  4.70 & 0.030 & $ 3.77 \times 10^{-11} $ & 14  \\ 
$G$ & 6230 & 4.74 & 0.003 & $ 3.24 \times 10^{-11} $ & 4.62 & 0.003 & $ 3.59 \times 10^{-11} $ & 6  \\ 
$m6300$ & 6300 & & & &  4.72 & 0.016 & $ 3.65 \times 10^{-11} $ & 13  \\ 
$j63$ & 6349 & 4.55 & 0.010 & $ 3.45 \times 10^{-11} $ & 4.49 & 0.010 & $ 3.66 \times 10^{-11} $ & 2  \\ 
$S_V$ & 6534 & & & &  4.28 &  & $ 3.62 \times 10^{-11} $ & 11  \\ 
$R_J$ & 6695 & 4.39 &  & $ 3.30 \times 10^{-11} $ & 4.34 &  & $ 3.46 \times 10^{-11} $ & 7  \\ 
$m6710$ & 6710 & & & &  4.68 & 0.016 & $ 3.34 \times 10^{-11} $ & 13  \\ 
$m7100$ & 7100 & & & &  4.66 & 0.016 & $ 3.03 \times 10^{-11} $ & 13  \\ 
$R_4$ & 7166 & & & &  4.34 &  & $ 3.11 \times 10^{-11} $ & 15  \\ 
$j72$ & 7222 & 4.37 & 0.010 & $ 2.79 \times 10^{-11} $ & 4.33 & 0.010 & $ 2.87 \times 10^{-11} $ & 2  \\ 
$m7400$ & 7400 & & & &  4.60 & 0.016 & $ 2.95 \times 10^{-11} $ & 13  \\ 
$G_{RP}$ & 7726 & 4.23 & 0.005 & $ 2.67 \times 10^{-11} $ & 4.18 & 0.004 & $ 2.79 \times 10^{-11} $ & 6  \\ 
$j80$ & 7993 & 4.18 & 0.010 & $ 2.40 \times 10^{-11} $ & 4.17 & 0.010 & $ 2.42 \times 10^{-11} $ & 2  \\ 
$I_J$ & 8565 & 4.04 &  & $ 2.26 \times 10^{-11} $ & 4.01 &  & $ 2.32 \times 10^{-11} $ & 7  \\ 
$j86$ & 8577 & 4.12 & 0.010 & $ 2.05 \times 10^{-11} $ & 4.11 & 0.010 & $ 2.07 \times 10^{-11} $ & 2  \\ 
$j99$ & 9813 & 4.04 & 0.010 & $ 1.64 \times 10^{-11} $ & 4.00 & 0.010 & $ 1.69 \times 10^{-11} $ & 2  \\ 
$j110$ & 11037 & 3.91 & 0.010 & $ 1.24 \times 10^{-11} $ & 3.88 & 0.010 & $ 1.28 \times 10^{-11} $ & 2  \\ 
$J$ & 12350 & 3.63 & 0.292 & $ 1.11 \times 10^{-11} $ & 3.78 & 0.254 & $ 9.62 \times 10^{-12} $ & 8  \\ 
$H$ & 16620 & 3.37 & 0.218 & $ 5.10 \times 10^{-12} $ & 3.40 & 0.216 & $ 4.94 \times 10^{-12} $ & 8  \\ 
$K_S$ & 21590 & 3.26 & 0.286 & $ 2.13 \times 10^{-12} $ & 3.24 & 0.244 & $ 2.17 \times 10^{-12} $ & 8  \\ 
$W1$ & 33526 & 3.37 & 0.132 & $ 3.66 \times 10^{-13} $ & 3.20 & 0.076 & $ 4.29 \times 10^{-13} $ & 9  \\ 
$W2$ & 46028 & 2.70 & 0.053 & $ 2.01 \times 10^{-13} $ & 2.47 & 0.071 & $ 2.49 \times 10^{-13} $ & 9  \\ 
$S9W$ & 82283 & 3.11 & 0.004 & $ 1.20 \times 10^{-14} $ & 2.95 & 0.011 & $ 1.39 \times 10^{-14} $ & 10  \\ 
$W3$ & 115608 & 3.34 & 0.014 & $ 3.00 \times 10^{-15} $ & 3.08 & 0.010 & $ 3.81 \times 10^{-15} $ & 9  \\ 
$L18W$ & 176094 & 3.08 & 0.042 & $ 6.06 \times 10^{-16} $ & 2.87 & 0.070 & $ 7.40 \times 10^{-16} $ & 10  \\ 
$W4$ & 220883 & 3.24 & 0.021 & $ 2.58 \times 10^{-16} $ & 3.11 & 0.018 & $ 2.91 \times 10^{-16} $ & 9  \\ 
\enddata
\tablecomments{1. \cite{thompsondata}, 2. \cite{johnsondata}, 3. \cite{paunzendata}, 4. \cite{moreldata}, 5. \cite{tycho2}, 6. \cite{gaiaedr3}, 7. \cite{mermillioddata}, 8. \cite{2mass}, 9. \cite{wise}, 10. \cite{akari}, 11. \cite{vilnius}, 12. \cite{geneva}, 13. \cite{clampitt}, 14. \cite{sdss}, 15. \cite{wbvrphoto}  }
\end{deluxetable*}

\clearpage 


\bibliography{paper}{}
\bibliographystyle{aasjournal}


\end{document}